\documentclass[useAMS,usenatbib,onecolumn]{mn2e}
\usepackage{graphicx}
\usepackage{bm}
\usepackage{epsfig}
\usepackage{amssymb, ulem}
\usepackage{amsmath}
\usepackage{multirow,booktabs}
\usepackage{euscript}
\usepackage{empheq}
\usepackage{amsfonts}
\usepackage{hyperref}
\usepackage{cleveref}
\usepackage{courier}
\usepackage[utf8]{inputenc}
\usepackage[T1]{fontenc}
\usepackage{lipsum}
\usepackage{mathtools}
\usepackage{cuted}
\usepackage{bm}
\usepackage{booktabs}
\usepackage{tabularx}
\usepackage[makeroom]{cancel}
\usepackage{cancel}
\usepackage{float}
\usepackage{placeins}
\usepackage{newunicodechar}

\usepackage[dvipsnames]{xcolor}

\hypersetup{
    colorlinks,
    linkcolor={red!80!black},
    citecolor={blue!80!black},
    urlcolor={RubineRed}
}

\newcommand{\bea}{\begin{eqnarray}}
\newcommand{\eea}{\end{eqnarray}}
\newcommand{\beq}{\begin{equation}}
\newcommand{\eeq}{\end{equation}}

\newcommand{\simless}[0]{\mathbin{\lower 3pt\hbox
   {$\rlap{\raise 5pt\hbox{$\char'074$}}\mathchar"7218$}}}
\newcommand{\simgreat}[0]{\mathbin{\lower 3pt\hbox
   {$\rlap{\raise 5pt\hbox{$\char'076$}}\mathchar"7218$}}}

\newcommand{\figref}[1]{figure \ref{#1}}

\newcommand{\capfigref}[1]{Figure \ref{#1}}

\newcommand{\eqnref}[1]{eq. (\ref{#1})} 
\newcommand{\eqnrefs}[1]{eqs. (\ref{#1})} 
\newcommand{\eqnrefbare}[1]{(\ref{#1})} 
\newcommand{\capeqnref}[1]{Equation (\ref{#1})}
\newcommand{\capeqnrefs}[1]{Equations (\ref{#1})}

\graphicspath{{}{figures/}}

\title[One-point PDF from spherical collapse.]{One-point probability distribution function (PDF) from spherical collapse: Early Dark Energy (EDE) vs. $\Lambda$CDM}
\author[Ankush Mandal and Sharvari Nadkarni-Ghosh]{Ankush Mandal$^{1,2}$\thanks{E-mail:ankushm@iucaa.in} and Sharvari Nadkarni-Ghosh$^{1}$\thanks{E-mail:nsharvari@gmail.com, sharvari@iitk.ac.in} \\
$^{1}$Department of Physics, I.I.T. Kanpur, Kanpur, U.P. 208016 India  \\
$^{2}$Inter-University Centre for Astronomy and Astrophysics, Pune-411007, India\\
}
\begin{document}
\date{}
\pagerange{\pageref{firstpage}--\pageref{lastpage}} \pubyear{}

\maketitle
\label{firstpage}

\begin{abstract}

We compute the one-point PDF of an initially Gaussian dark matter density field using spherical collapse (SC). We compare the results to other forms available in the literature and also compare the PDFs in the $\Lambda$CDM model with an early dark energy (EDE) model. We find that the skewed log-normal distribution provides the best fit to the non-linear PDF from SC for both cosmologies, from $a=0.1$ to 1 and for scales characterized by the comoving width of the Gaussian: $\sigma_G = 0.5, 1, 2$. To elucidate the effect of cosmology, we examine the linear and non-linear growth rates through test cases. For overdensities, when the two models have the same initial density contrast, the differences due to cosmology are amplified in the non-linear regime, whereas, if the two models have the same linear density contrast today, then the differences in cosmology are damped in the non-linear regime. This behaviour is in contrast with voids, where the non-linear growth becomes `self-regulatory' and is less sensitive to cosmology and initial conditions. To compare the PDFs, we examine the difference of the PDFs and evolution of the width of the PDF. The trends with scale and redshift are as expected. A tertiary aim of this paper was to check if the fitting form for the non-linear density-velocity divergence relation, derived for constant equation of state ($w$) models by Nadkarni-Ghosh holds for the EDE model. We find that it does with an accuracy of 4\%, thus increasing its range of validity.

\end{abstract}

\begin{keywords}
cosmology: theory - large-scale structure of Universe - dark energy
\end{keywords}

\section{Introduction}
\label{sec:intro}

The large scale structure observed today is believed to have grown from tiny fluctuations laid down during the inflationary era in the history of the Universe. Observations of the cosmic microwave background suggest that the one-point probability distribution function (PDF) of these primordial fluctuations is very close to a Gaussian (Planck Collaboration\footnote{https://www.cosmos.esa.int/web/planck, Planck 2018 results. IX.}). At early epochs, when the evolution is linear, the shape of the PDF remains Gaussian and hence symmetric about the origin. However, this symmetry is broken in the non-linear regime. For instance, the density contrast in overdense virialized regions is of the order of a hundred, but it never drops below -1 in void regions, since the total density is always positive. 
Tracking this (secondary) non-Gaussianity is important because the shape of the PDF carries information about the evolution history and hence can be used to constrain cosmological parameters. While the two-point and higher-order correlation functions give information about the spectral distribution, there is still observational interest in additionally measuring the matter density PDF, either directly in terms of the galaxy counts-in-cell distribution or indirectly through the lensing convergence \citep{bel_vimos_2016, liu_cmb_2016, patton_cosmological_2017, hurtado-gil_best_2017,clerkin_testing_2017,gruen_density_2018,friedrich_density_2018}. This has been shown to provide complementary constraints and help break degeneracies in parameter estimation.

The history of the one-point PDF begins almost a century ago, when \cite{hubble_distribution_1934} examined the frequency distribution of about 44000 galaxies distributed 
over three-fourth of the sky and found it to be close to log-normal. Following this, there were other many other statistical measures proposed to quantify the departure from Gaussianity, including the extensive work by \citet{neyman_alternative_1962} and \citet{scott_distribution_1962}. A detailed historical summary can be found in \cite{peebles_large-scale_1980}. The early 80s and 90s saw the advent of phenomenological models some of which assumed that gravitational clustering can be described by equilibrium thermodynamics \citep{saslaw_thermodynamics_1984,suto_gravitational_1990,lahav_non-gaussian_1993}. Another phenomenological model which has remained popular till date is the log-normal model proposed by \cite{coles_lognormal_1991}. This model is partly motivated by the theoretical fact that the full non-linear continuity equation combined with linear statistics for the velocity field, predicts a log-normal statistics for the density field. This form has proven to be a good fit in the quasi-linear regime both theoretically \citep*{kofman_cosmological_1993, coles_testing_1993, ohta_evolution_2003} and observationally \citep{colombi_experimental_2000,szapudi_experimental_2000,wild_2df_2005,clerkin_testing_2017}. It has also been proposed to fit the weak lensing convergence field \citealt{taruya_lognormal_2002,hilbert_cosmic_2011}) as well as to predict the void distribution function \citep{russell_log-normal_2017}. However, its validity in the non-linear regime is less established (for e.g., \citealt{colombi_skewed_1994,bernardeau_properties_1995, ueda_counts--cells_1996, szapudi_recovering_2004, repp_precision_2018}). Recently, \cite{xavier_improving_2016} have also pointed out a mathematical inconsistency in simultaneous modelling the joint distribution of the density and convergence fields and have proposed modifications to the standard form. Numerical simulations have also shown deviations from this form and alternate fitting forms and schemes have been proposed (for e.g., \citealt{shin_new_2017,klypin_density_2018,tosone_beyond_2020}).

On the theoretical side, analytical modelling of the one-point PDF has been of interest for over three decades. In the 90s, Kofman and Bernardeau and their collaborators used  perturbation theory (PT) in the Eulerian and Lagrangian frames to derive the density PDFs and its higher-order moments \citep{kofman_cosmological_1993,kofman_evolution_1994,bernardeau_effects_1994,bernardeau_skewness_1994,bernardeau_properties_1995,colombi_extended_1997}. Extensions of this work were done by \cite{fosalba_cosmological_1998} who simplified the higher-order perturbative terms using spherical dynamics so as to apply the formalism to non-Gaussian initial conditions. \cite{juszkiewicz_weakly_1995} and \cite{gaztanaga_gravitational_2000} combined PT with the Edgeworth expansion to get better approximations in the non-linear regime. \cite{taylor_evolution_2000} combined second-order PT with the Chapman-Kolmogorov equation to understand the evolution of the distribution. 
Estimates based on local dynamics such as spherical and ellipsoidal collapse were given by 
\citet{ohta_evolution_2003,ohta_cosmological_2004} who formulated the differential equations for the evolution of the one-point PDFs as well as \citet{lam_ellipsoidal_2008, lam_perturbation_2008} who used excursion sets to construct the PDF in both real and redshift space. 

One non-perturbative approach was put forth by Valageas who recast the solution to the collisionless Boltzmann equation as a path integral to be solved within the saddle-point approximation using the method of steepest descent. The dynamics near the saddle point depends only on the scale separation because of the homogeneity and isotropy of the Universe and spherical symmetry is invoked to provide a unique mapping between the initial linear state and the final non-linear state of the system \citep{valageas_structure_1998,valageas_dynamics_2001,valageas_dynamics_2002-1,valageas_dynamics_2002-3,valageas_dynamics_2002,valageas_dynamics_2002-2,valageas_evolution_2004}.  This was further developed and applied to compute the statistic of density profiles by \citet{bernardeau_statistics_2014,bernardeau_joint_2015}. Very recently, \cite{ivanov_non-perturbative_2019} have extended this calculation to compute the effect of fluctuations around the spherical saddle point, which they argue is necessary for the consistency of the theory. 
\citet{bernardeau_large_2016} have recently proposed another non-perturbative approach that uses Large Deviations Theory to compute the cumulant generating functions. This approach also relies on having a unique mapping between initial and final densities given by local approximations like spherical or cylindrical collapse \citep{uhlemann_back_2016,uhlemann_cylinders_2018}. This method has been applied to a variety of problems: to compute marginal and joint density and velocity-divergence PDFs \citep{uhlemann_two_2017}, the weak lensing convergence PDF \citep{reimberg_large_2018, barthelemy_nulling_2020} and to constrain dark energy models \citep{codis_encircling_2016}. 
																																																																	
Many of the above theoretical methods use spherical collapse dynamics as a proxy for the non-linear regime. The main advantage is that the equations are simple second-order ordinary differential equations (ODEs). Analytic solutions are available for pure matter cosmologies and numerical solutions can be easily generated not only for $\Lambda$CDM but also for many other non-standard cosmological models such as those with a constant equation of state ($w \neq -1$) or those with a time-varying $w$. The exceptions are cases where dark matter and dark energy are coupled or Einstein's law of gravity is modified since these systems are higher-order with additional perturbation variables. In spite of this simplicity, most of the aforementioned investigations that invoke spherical dynamics have been restricted to pure matter or $\Lambda$CDM only. 

In this paper, we use the spherical collapse model to compute the one-point PDF and compare the case of an Early Dark Energy (EDE) model with the usual $\Lambda$CDM cosmology. The physical assumption is that each spatial point evolves independently (local approximation) as if it were an isolated spherical perturbation in an expanding background cosmology. The stochasticity is in the initial conditions: they are drawn from a random Gaussian distribution. This paper has two primary aims. First, we want to check how well the non-linear PDF from spherical collapse compares with other fitting forms in the literature and second, we want to understand the effect of early dark energy on structure, in particular on the one-point PDF. \citet{nadkarni-ghosh_non-linear_2013} also considered spherical collapse to obtain the density-velocity divergence relation (DVDR) in the non-linear regime. However, the investigation was restricted to dark energy models with constant equation of state ($w=w_0$). A third, tertiary aim of this work is to also check if the fitting form for the non-linear DVDR obtained in N13 also works for the EDE model with a time-varying $w$.

The paper is organized as follows. \S \ref{sec:setup} sets up the equations for non-linear evolution of the density field for spherically symmetric perturbations expanding in a background cosmology consisting of a dark energy component that neither clusters nor couples to the matter component. \S \ref{sec:growthrate} discusses the growth rates in the linear and non-linear regime and compares the behaviour of the EDE model with the standard $\Lambda$CDM model. Here we accomplish our third aim of checking how well the non-linear DVDR relation obtained in N13 works for the EDE model by comparing the theoretical and numerical non-linear growth rates. 
\S \ref{sec:PDF} gives the details of the numerical runs performed to generate the PDF and compares the results to fitting forms existing in the literature. We consider three forms: the theoretical prediction based on perturbation theory \citep{bernardeau_effects_1994}, the skewed log-normal \citep{colombi_skewed_1994} and the generalized normal distribution, version 2 \citep{shin_new_2017}. These estimates are in the Eulerian frame, whereas, the PDFs obtained through spherical collapse are Lagrangian PDFs and we discuss the transformation between the two. \S \ref{sec:comp} compares the PDFs for $\Lambda$CDM and EDE models in terms of (a) the difference of the PDFs and the (b) evolution of the width of the PDF. We conclude in \S \ref{sec:conclusion}.
%
%
\section{The physical set-up}
\label{sec:setup}
\subsection{Non-linear evolution equations in the Spherical Collapse (SC) model}
The physical system is a constant density (top-hat) sphere, evolving in a homogenous and isotropic, flat background universe consisting of pressureless matter and dark energy with equation of state $w$, which may vary with time. Matter is allowed to cluster whereas dark energy is assumed to be uniform, and non-interacting with the matter sector. For such cosmologies, the scale factor $a$ of the background evolves as 
 \beq 
H^2= H_0^2 \left[ \frac{\Omega_{m,0}a_0^3}{a^3} + \Omega_{de,0} \exp\left\{- 3 \int_{a_0}^a (1+ w(y))\frac{dy}{y} \right\} \right],
 \label{backeqn}
 \eeq
where $H={\dot a}/a$. The `dot' denotes derivative with respect to coordinate time, $y$ is the dummy variable used for the integral and $a_0$ is the value of the scale factor today, usually set to 1. The matter and dark energy parameters are denoted as $\Omega_m$ and $\Omega_{de}$ respectively and the subscript `0' denotes their values today.
Using the conservation of energy density, the density parameters at any epoch are related to their values today as
\bea
 \Omega_{m}(a) & =& \frac{H_0^2 a_0^3\Omega_{m,0} }{ H^2a^3} \\
\Omega_{de}(a)& = &\frac{H_0^2}{H^2}\Omega_{de,0} \exp\left\{- 3 \int_{a_0}^a (1+ w(y))\frac{d y}{y}\right\}
\eea
For a flat universe, $\Omega_m(a) + \Omega_{de}(a) =1$ for all epochs. Let the origin be at the centre of the spherical perturbation and let ${\bf r}$ and ${\bf x}$ be the physical and comoving radius of the outer edge.  The perturbed sphere may be surrounded by a compensating region, which is either a mass shell or a void, depending upon whether the perturbation is underdense or overdense with respect to the background. Let ${\bar \rho}_m$ and $\rho_m$ be the density of the background and perturbation respectively. 
For the dark energy models considered in this paper, an initial top-hat stays a top-hat and the dynamics can be solely described by two purely time-dependent quantities: fractional overdensity ($\delta$) and the scaled divergence of the peculiar velocity ($\Theta$) defined as 
\bea 
\delta &=& \frac{\rho_m}{{\bar \rho}_m} -1, \\
\Theta &=& \frac{1}{H} \nabla_r \cdot {\bf v}, 
\eea
where ${\bar \rho}_m$ and $\rho_m$ are the densities of the background and perturbation respectively and ${\bf v}  = {\dot {\bf r}} - H {\bf r} = a {\dot {\bf x}}$ is the peculiar velocity.
For a pure matter cosmology, it is possible to invoke Birkoff's theorem and treat the perturbed sphere as a separate universe whose evolution is dictated by an equation analogous to \eqnref{backeqn}, with a different scale factor and density parameter. However, extensions to dark energy cosmologies are not obvious. Here we adopt the approach of starting with the hydrodynamic equations: the generalized continuity and Euler equations along with the Poisson equation for the gravitational potential \citep{lima_newtonian_1997,abramo_physical_2009,pace_spherical_2010,pace_implementation_2017}. These are valid for general non-interacting dark energy cosmologies, in the sub-horizon limit and in the absence of shell-crossing. The resulting equations for $\delta$ and $\Theta$ are 
\bea
\label{deltaeq1}\delta' &=& -(1+\delta) \Theta,\\
\label{Thetaeq1}\Theta' &=& -\frac{3}{2} \Omega_m(a) \delta - \Theta \left(2 + \frac{H'}{H} \right) - \frac{\Theta^2}{3}.
\eea
The prime denotes derivative with respect to $\ln a$. The details of the derivation are in appendix \ref{app:eqns}. Combining \eqnref{deltaeq1} and \eqnref{Thetaeq1} gives one second order equation for the non-linear evolution of $\delta$. 
\beq
\delta'' + \left(2 + \frac{H'}{H}\right) \delta' - \frac{4}{3} \frac{\delta'^2}{(1+\delta)} - \frac{3}{2} \Omega_m(a) \delta(1+\delta)=0.
\label{deltaNLeqn}
\eeq
For the flat dark energy models, with varying equation of state,  \eqnref{deltaNLeqn} reduces to 
\beq
\delta'' + \frac{1}{2} \left[1-3w(a) \Omega_{de}(a)\right]\delta' - \frac{4}{3} \frac{\delta'^2}{(1+\delta)}  - \frac{3}{2} \Omega_m(a) \delta(1+\delta)=0.
\label{deltaNLeqn2}
\eeq
This system is also valid if the density has a radial variation: in that case $\delta$ has to be replaced by its spherically averaged value (N13).


\subsection{Initial conditions}
\capeqnref{deltaNLeqn2} can be completely solved by specifying $\delta$ and $\delta'$ at some initial time $a_{i}$. From \eqnref{deltaeq1}, we have $\delta'_{i} =- (1+\delta_{i}) \Theta_{i}$. In principle, $\delta_i$ and $\Theta_i$ are independent however, usually cosmological initial conditions assume the Zeldovich approximation \citep{zeldovich_gravitational_1970}. This states that the velocity field is proportional to the acceleration field; the proportionality constant is chosen so as to ensure that there are no perturbations at the big bang time. In linear, Eulerian perturbation theory, for the EdS cosmology, this is equivalent to setting the coefficient of the decaying mode to zero and the growing mode solution is $\delta(a) \propto a$, which gives $\delta'=\delta$ or $\Theta = -\delta$. In this paper, we use the generalization given by  \cite{linder_cosmic_2005} for dark energy cosmologies: $\Theta_i = \Omega_{m,i}^{0.55} \delta_i$. For sufficiently early epochs and sufficiently small $\delta_i$, this reduces to $\delta'_i =\delta_i$.  We will discuss extensions of this linear relation in \S \ref{sec:growthrate}. 

\subsection{Cosmological models}

In this paper, we primarily consider two models: (a) $\Lambda$CDM with $\Omega_{m,0}=0.29$, $\Omega_{de,0}=0.71$ and (b) Early Dark Energy (EDE hereafter) model with varying equation of state
\begin{equation}
w(a)=\frac{w_0}{(1-b \ln a)^2}
\label{eosEDE}
\end{equation}
with
\begin{equation}
b=-\frac{3w_0}{\text{ln}\Big(\frac{1-\Omega_{de,e}}{\Omega_{de,e}}\Big)+\text{ln}\Big(\frac{1-\Omega_{m,0}}{\Omega_{m,0}}\Big)},
\end{equation}
where $\Omega_{de,e}$ is the value of $\Omega_{de}$ at very early time ($a\rightarrow 0$), $w_0$ is the value of equation of state parameter at present day. This three-parameter model was first proposed by \cite{wetterich_phenomenological_2004} and later also considered by \cite{doran_early_2006}. The EDE model contains a low but non-vanishing amount of dark energy at early epoch. It can be 
characterized by the quantity
\beq 
{\bar \Omega_{de,sf} } = (\ln a_{tr}-\ln a_{eq})^{-1}\int_{\ln a_{eq}}^{\ln a_{tr}}\Omega_{de}(a)d\ln a, 
\label{omegasf}
\eeq
where $a_{eq}$ is the epoch of matter radiation equality and $a_{tr}$ is the epoch where dark energy starts to dominate ($a_{tr}\sim 1/3$). ${\bar \Omega_{de,sf} }$ denotes the average fraction of dark energy during the structure formation/matter domination era and using a combination of data sets it is constrained to be ${\bar \Omega_{de,sf} } \leq 0.04$  \citep*{doran_structure_2001,doran_impact_2007}. This constraint was imposed by \cite*{bartelmann_non-linear_2006} as well as \cite{grossi_impact_2008} who investigated  the impact of EDE on non-linear structure. They considered two EDE models with ${\bar \Omega_{de,sf} } =0.04$, but other parameters randomly picked from a Monte Carlo chain. The two EDE models did not give qualitatively different results, so, we choose one of them: $\Omega_{de,e} = 8 \times 10^{-4} $ and $w_0 =-0.99$.  The value for $\Omega_{m,0}$ was chosen to be the same as the $\Lambda$CDM model and the requirement of spatial flatness sets $\Omega_{de,0}$. An alternative approach is to consider the best fit values for an EDE cosmology, which in general will be different than those in $\Lambda$CDM \citep{jennings_simulations_2010}, but we do not adopt this approach here. We do not consider a larger range of EDE models because the main aim is to qualitatively contrast the growth of the PDF in $\Lambda$CDM and EDE scenarios.

Another commonly used parameterization for dark energy is the two-parameter CPL parameterization \citep{chevallier_accelerating_2001,linder_exploring_2003}
\beq
    w_{CPL}(a)=w_{0}+w_a(1-a), 
\label{eosCPL}
\eeq
where $w_0$ and $w_a$ are two free parameters to be set by data. This parameterization corresponds to a first order Taylor expansion of the equation of state around $a=1$, with $w_0$ being the value at $a=1$. Performing a similar expansion on \eqnref{eosEDE} gives 
\beq
w_{EDE}(a) = w_0  -2b w_0(1-a).
\label{EDEexp}
\eeq
Matching \eqnrefs{eosCPL} and \eqnrefbare{EDEexp} gives $w_a = -2 bw_0$. Most recent observational constraint from \citet{Plank_collab2018} is $w_0=-0.957$ and $w_a=-0.29$ combining the \textit{Planck}, SNe and BAO data. A more detailed comparison of the two models can be found in appendix \ref{app:EDECPLcompare}.

\capfigref{omega} shows the evolution of the matter density for the models considered in this paper.  For pedagogical reasons, we also include here and in the next section the EdS  (flat, pure matter) cosmology with $\Omega_m = 1$. The left panel shows the density parameters $\Omega_m$ and $\Omega_{de}$ in all three cosmologies. The EDE cosmology has a non-negligible amount of dark energy at early times. Assuming flatness, this amounts to a lower value of $ \Omega_m$ at early epochs. The EdS cosmology has $\Omega_m=1$ at all epochs. The right panel shows the ratio of the matter density parameters for the dark energy cosmologies: $ \Omega_{m, {\rm EDE}}/\Omega_{m, \Lambda {\rm CDM}}$. The ratio is minimum at around $a\sim 0.4$. The difference in growth rates between the two models is sensitive to this ratio; see \figref{fig:linrate}. 

\begin{figure}
\includegraphics[width=8cm]{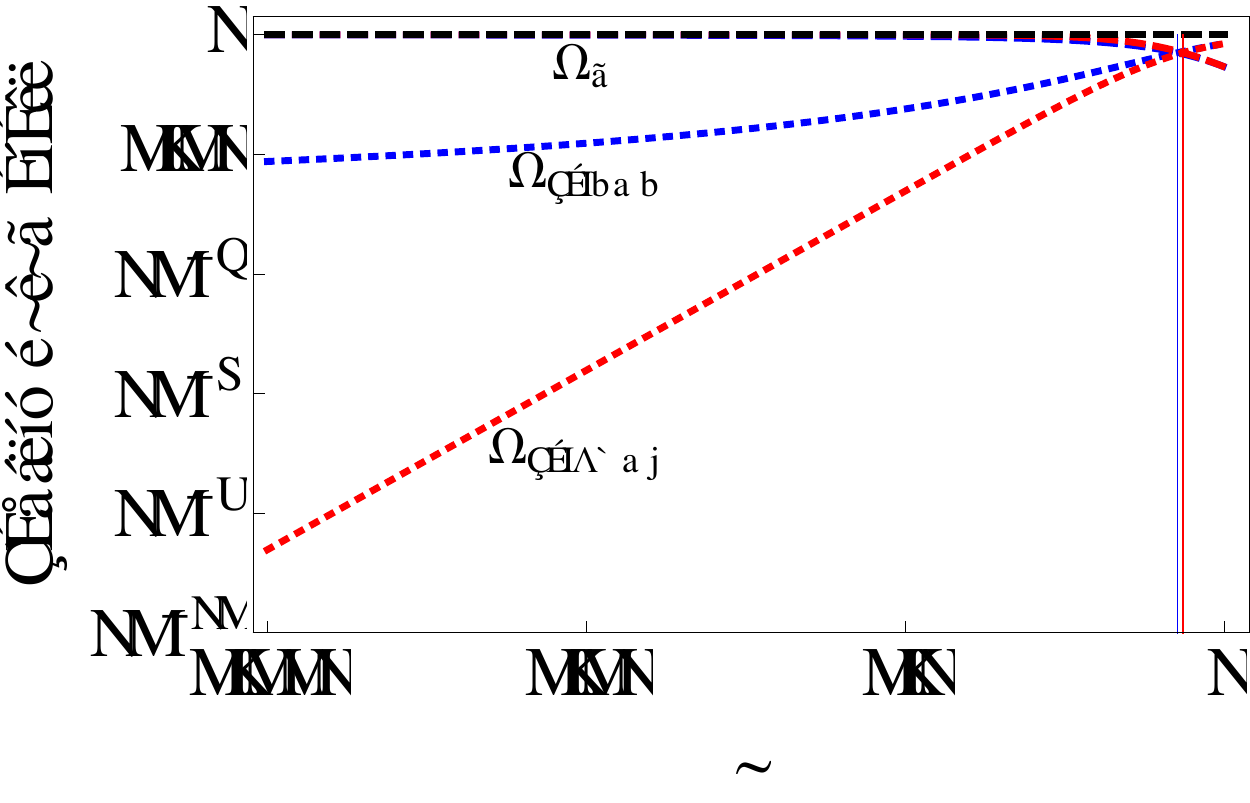}
\hspace{1cm}
\includegraphics[width=8cm]{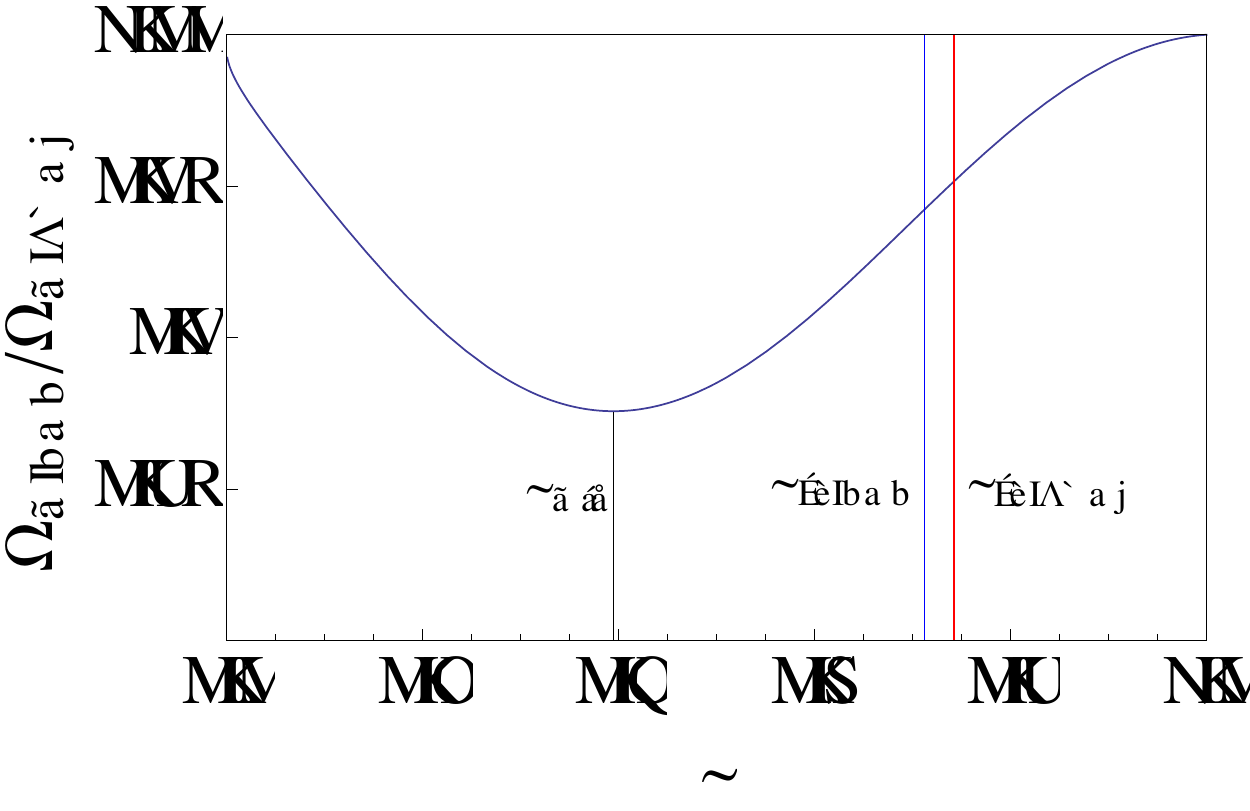}
\caption{Evolution of density parameters. Left panel: the matter density parameter is unity for the EdS cosmology. The EDE model has non-negligible dark energy at early epochs as compared to $\Lambda$CDM, but the $\Omega_m$ is still close to unity for both models. Right panel: the ratio of the matter density parameters in EDE relative to the $\Lambda$CDM cosmology. The ratio is minimum around $a\sim 0.4$. The vertical blue and red lines in both panels indicate the epoch of matter-dark energy equality for EDE and $\Lambda$CDM respectively. They are very close and almost indistinguishable on a log-axis.}
\label{omega}
\end{figure}


\section{Linear and non-linear growth rates}
\label{sec:growthrate}
The non-linear evolution of $\delta$ in the SC model is given by \eqnref{deltaNLeqn2}. The growth rate $f$ is defined as 
\beq
f= \frac{d \ln \delta}{d \ln a}. 
\label{growthrate}
\eeq
This definition is valid in the linear as well as non-linear regime.  In the linear regime, $\delta \propto D(a)$, where $D(a)$ is the growth factor (see appendix \ref{app:eqns}) and $f = D'(a)/D(a)$. Using the continuity equation (\eqnref{deltaeq1}), $f$ can also be expressed as
\beq 
 f = -\frac{1+\delta}{\delta} \Theta. 
\label{nonlinrate}
\eeq
In non-linear regime, $f$ is a function of the instantaneous $\delta$. The higher the $\delta$, the higher the growth rate. 
Knowing the relation between the $\delta$ and $\Theta$, it is possible to compute the growth rate as a function of density. 
This relation has been studied extensively by various authors in the past, using perturbation theory (e.g., \citealt{bernardeau_quasi-gaussian_1992}) or numerical simulations. 
\cite{bilicki_velocity-density_2008} gave a fit based on analytical solutions of SC. N13, investigated the $w$ dependence of this relation, also using SC, but without relying on the analytic solutions. The main idea was to impose the condition that there be `no perturbations at the big bang'. This is a generalization of the idea of `no decaying mode', which is the basis of the `Zeldovich approximation' \citep{zeldovich_gravitational_1970,peebles_large-scale_1980}. This is also sometimes referred to as the `slaving condition'. The assumption that the initial peculiar velocity field is `slaved' to the initial gravitational field has found various applications in the literature. For example, \cite{brenier_reconstruction_2003} combined this condition with the MAK reconstruction formalism to evolve perturbations back in time to obtain information about the primordial initial conditions. \cite{buchert_lagrangian_1992} showed that this condition was a special case of a more general first-order Lagrangian Perturbation Theory (LPT). \cite{nadkarni-ghosh_extending_2011}, while investigating the convergence properties of LPT, found that this condition, imposed in the SC model for an EdS cosmology, traces out a unique curve in the two dimensional $\delta-\Theta$ phase space. This curve, called the `Zeldovich curve' is an invariant set of the dynamics and this property was exploited to extend the range of validity of LPT. \cite{nadkarni-ghosh_non-linear_2013}\footnote{In N13, $\delta_v$ was used instead of $\Theta$; $\Theta = 3 \delta_v$.} extended the phase space analysis to include constant equation of state cosmologies and showed that the invariant property of the `Zeldovich curve' still holds -  perturbations that start with the slaving condition, maintain the condition and those that do not start out slaved, eventually get slaved. Thus, this condition acts like an attractor of the dynamical system defined by \eqnref{deltaeq1} and \eqnrefbare{Thetaeq1} and it corresponds to the late time (non-linear) density velocity-divergence relation (DVDR). The fitting form obtained in N13 based constant $w$ cosmologies is:

\beq
\label{Thetafit}
\Theta_{fit}(\Omega_m, w,\delta) = \left \{\begin{array}{cc}
3 A(\Omega_m, w) \left(1-(1+\delta)^{B(\Omega_m, w)}\right), & -1\leq \delta \leq 1\\ \\
3 \Omega_m^{\gamma_1(w) + \gamma_2(w)} \left[(1+\delta)^{1/6} -(1+\delta)^{1/2} \right], &   1\leq \delta \leq10.
\end{array}
\right.
\eeq
This fit is an extension of the forms given by \cite{bernardeau_quasi-gaussian_1992} and \cite{bilicki_velocity-density_2008}.  Substituting from \eqnref{Thetafit} gives 
\beq
f_{fit}(\Omega_m, w,\delta) = \left \{\begin{array}{cc}
-3 A(\Omega_m, w) \delta^{-1}\left((1+\delta)-(1+\delta)^{B(\Omega_m, w)+1}\right), & -1\leq \delta \leq 1\\ \\
-3 \Omega_m^{\gamma_1(w) + \gamma_2(w)}\delta^{-1} \left[(1+\delta)^{7/6} -(1+\delta)^{3/2} \right], &   1\leq \delta \leq10
\end{array}
\right.
\label{growthfit}
\eeq
where 
\beq
A(\Omega_m, w) = \frac{1}{2}\Omega_m^{\gamma_1(w)}, \; B(\Omega_m, w) = \frac{2}{3}\Omega_m^{\gamma_2(w)}, \; \gamma_1(w)  = 0.56(-w)^{-0.08}\;\; {\rm and}\;\;\gamma_2(w)= -0.01(-w)^{-1.18}.
\label{eqnAB}
\eeq
and  $\Omega_m$ is evaluated at the epoch of interest.
In the linear regime, the fitting function reduces to 
\beq
f_{lin,fit}(\Omega_m, w) = \Omega_m^{\gamma_1(w) + \gamma_2(w)}, 
\label{fitw}
\eeq
which for the $\Lambda$CDM case, agrees with the commonly used parameterization of \cite{linder_cosmic_2005} $f_{Linder} \approx \Omega_m^{0.55}$.

One of the aims of this paper is to check the validity of \eqnref{growthfit} for the EDE model. To do this, we compare this form given by \eqnref{growthfit} to the numerically computed growth factor obtained by solving \eqnref{deltaNLeqn2} for Zeldovich initial conditions. 
 We consider the following cases: 
\begin{itemize}
 \item Case (1a): Overdensity with same initial amplitude for all three cosmologies [$\delta(a=0.001) = 1.52 \times 10^{-3}$].
 \item Case (1b): Void with same initial amplitude for all three cosmologies [$\delta(a=0.001) = -1.52 \times 10^{-3}$].
 \item Case (2a):  Overdensity with same linear amplitude today for all three cosmologies [$\delta(a=1) = 1 $].
  \item Case (2b): Void with same "linear" amplitude today for all three cosmologies [$\delta(a=1) = -1$]. 
 \end{itemize} 
For cases (2a) and (2b), the initial values at $a_i=0.001$ are obtained by multiplying by the linear growth factor normalized to unity at $a=1$. This factor is denoted as ${\tilde D}(a)$ and expression for a general cosmology is in appendix \ref{app:eqns}. This gives $\delta_{i,EDE}= \pm 1.52 \times 10^{-3},\delta_{i,\Lambda CDM}= \pm 1.29 \times 10^{-3},\delta_{i,EdS}= \pm0.001$.
We note that the limit $\delta_L = -1$ does not correspond to the linear regime, but the generated initial amplitudes are well defined. The rationale behind considering two classes of initial conditions is the following. It is clear that the presence of early dark energy will reduce the growth rate of structure. The simplest way to illustrate this is to consider the same initial conditions for all three cosmologies. This corresponds to cases (1a) and (1b). However, observations constrain the linear power spectrum today i.e., at $a=1$. Thus, it is important to understand the differences in non-linear growth, subject to this constraint. This corresponds to cases (2a) and (2b).  By construction, the EDE solution is the same for classes (1) and (2). 
\begin{figure}
\includegraphics[width=7cm]{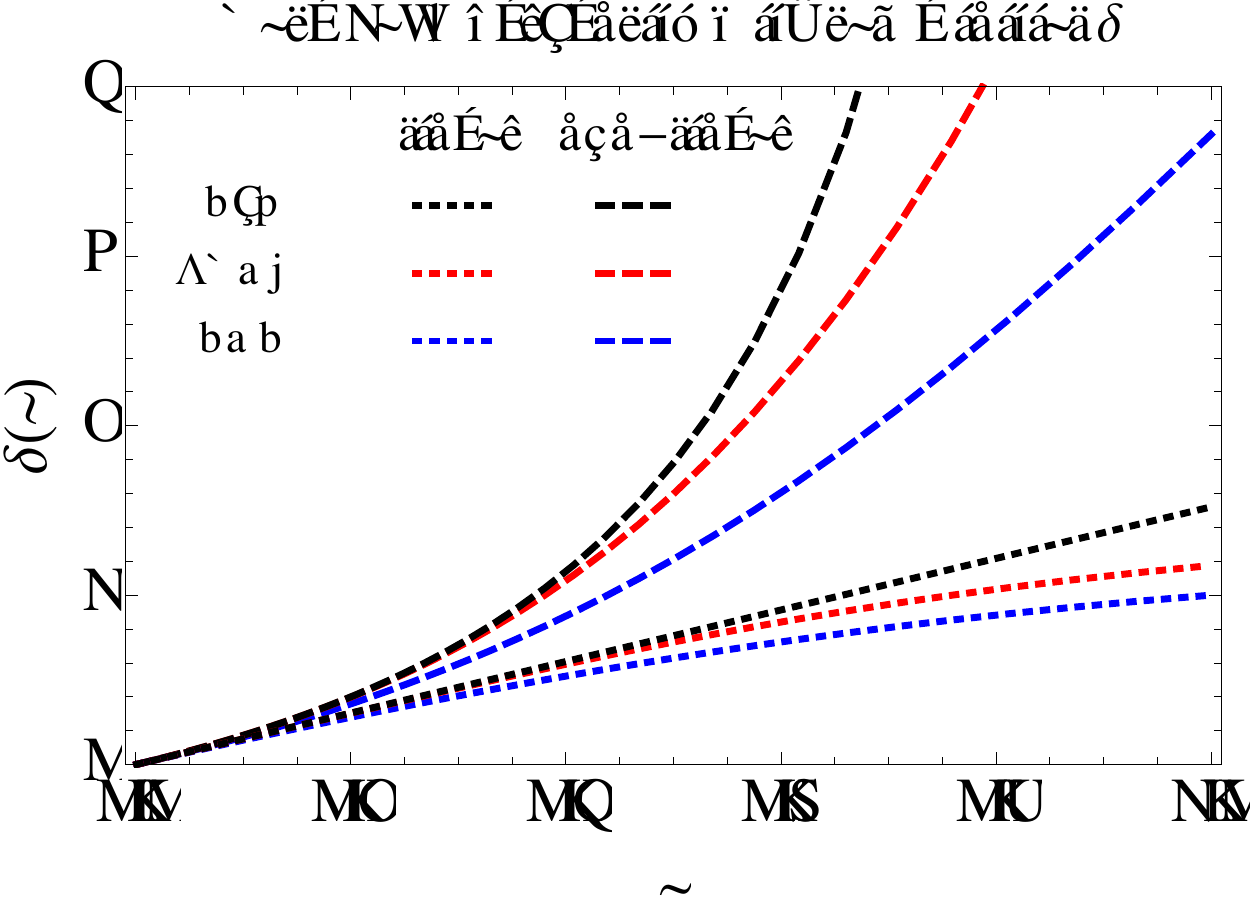} \hspace{1cm}
\includegraphics[width=7cm]{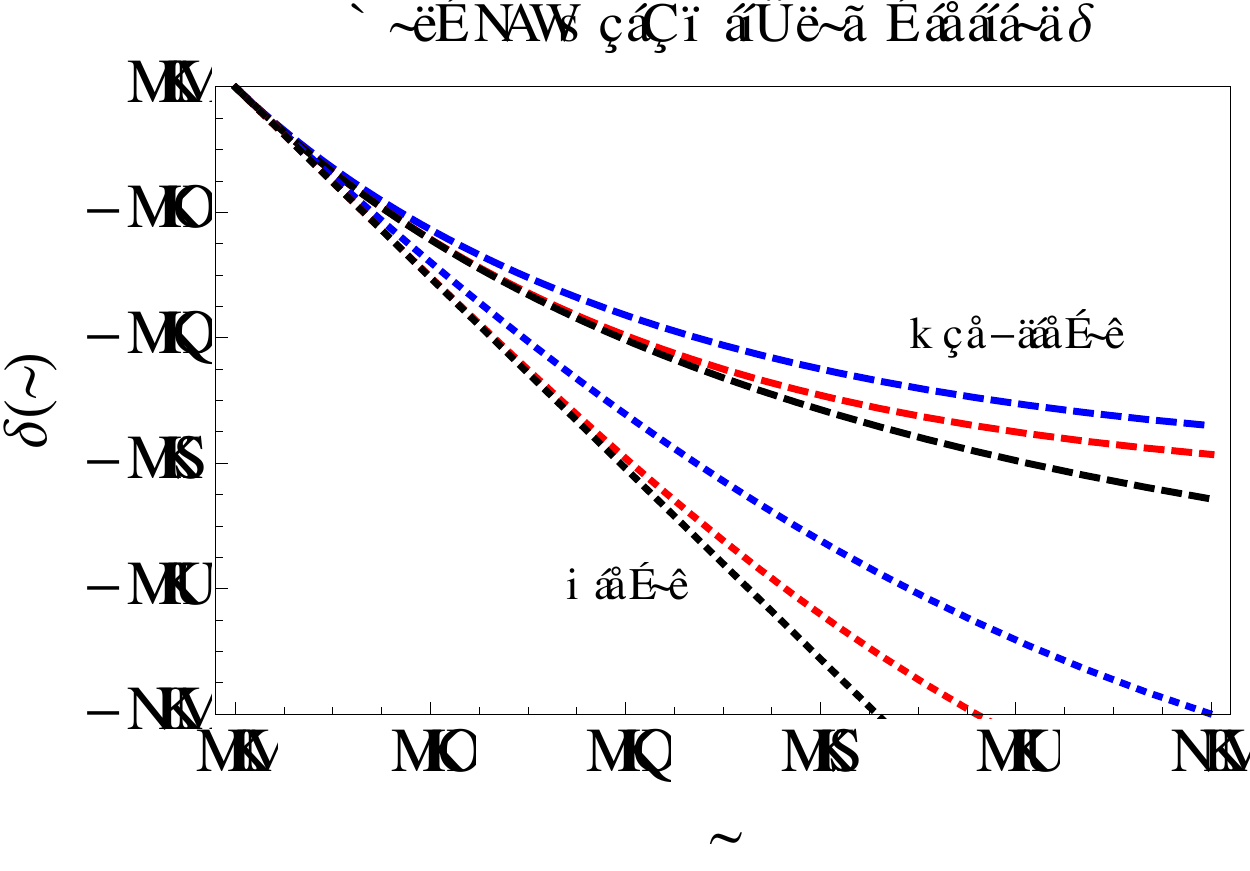}\\
\includegraphics[width=7cm]{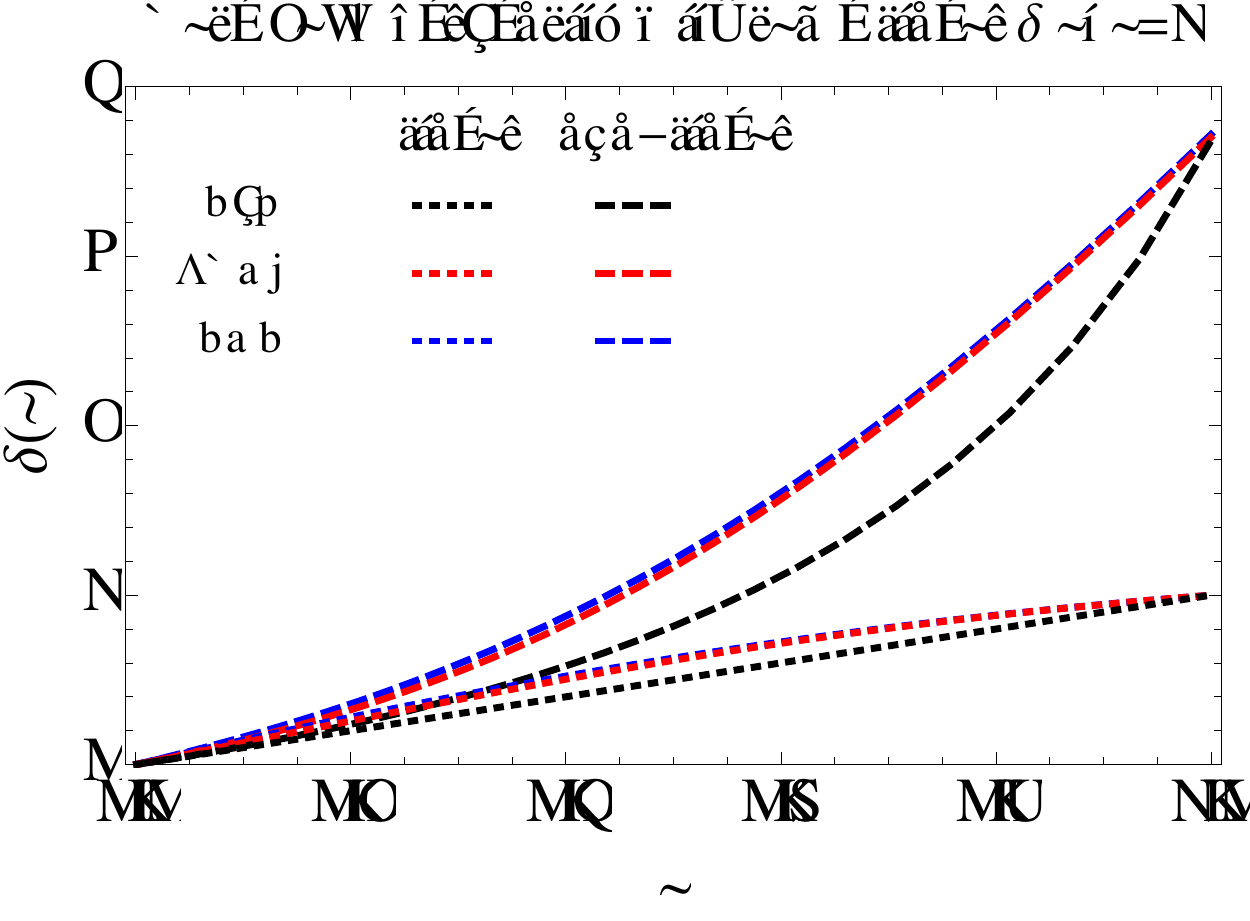} \hspace{1cm}
\includegraphics[width=7cm]{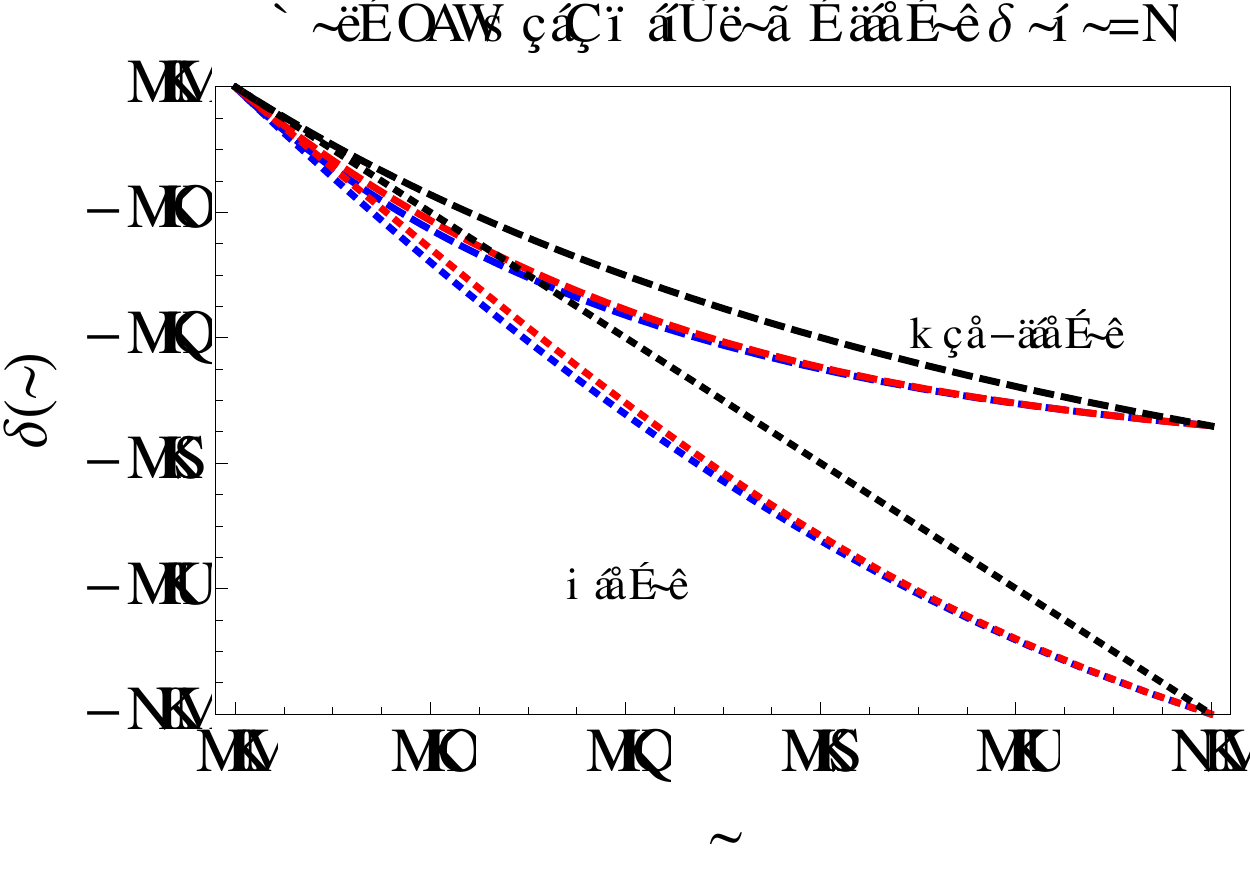}
\caption{Solutions for $\delta$ as a function of scale factor for the four cases defined in the text. (1a) and (1b) have same initial conditions for all three cosmologies whereas, (2a) and (2b) have same conditions at $a=1$. The dependence on cosmology is clear in cases (1a) and (1b) whereas, in cases (2a) and (2b), a lower $\Omega_m$ for the EDE model gets compensated by a higher initial amplitude and the deviations due to cosmology are less obvious. The dark energy models in both (2a) and (2b) are more non-linear than the pure matter cosmology.  
Note also that for voids, the non-linear solutions grows slower than in the linear regime. As a check, it is useful to note that in the pure matter (EdS) cosmology, $\delta \propto a$ in the linear regime. 
 }
\label{fig:solns}
\end{figure}

\capfigref{fig:solns} shows the evolution of $\delta$ as a function of $a$ for the four cases. The following features can be noted. First, consider the overdensities. The differences between cosmologies are very clear in case (1a), whereas, they are much lower in (2a); in fact, the solutions for the $\Lambda$CDM and EDE cosmologies, almost coincide. This is because in case (2a), the lower $\Omega_m$ in the EDE cosmology is compensated with a higher initial amplitude.
The effect is even more pronounced when comparing EdS and dark energy cases. The initial amplitude is low enough that the value of $\delta$ is always less than the dark energy cases; however, the growth rate is higher so the solution evolves to the same $\delta$ value at $a=1$. 
For voids, we first note that the linear evolution is `faster' than the non-linear evolution. This is because a void is underdense as compared to the background and hence grows slower. Mathematically, in the linear regime, $|\delta'| \propto |\Theta|$ whereas, in the non-linear regime $|\delta'| \propto |(1+\delta) \Theta|$. Since $(1+\delta) <1$, the non-linear growth is slower than the linear growth. Comparing cases (1a) and (1b), we note a similar effect as seen in overdensities. 
In case (1b), the lower value of $\Omega_m$ gives a slower evolution in EDE cosmology as compared to $\Lambda$CDM. This difference is reduced in case (2b) because the lower value of $\Omega_m$ is compensated by a higher initial amplitude (absolute value) in the EDE.  

\begin{figure}
\includegraphics[width=7cm]{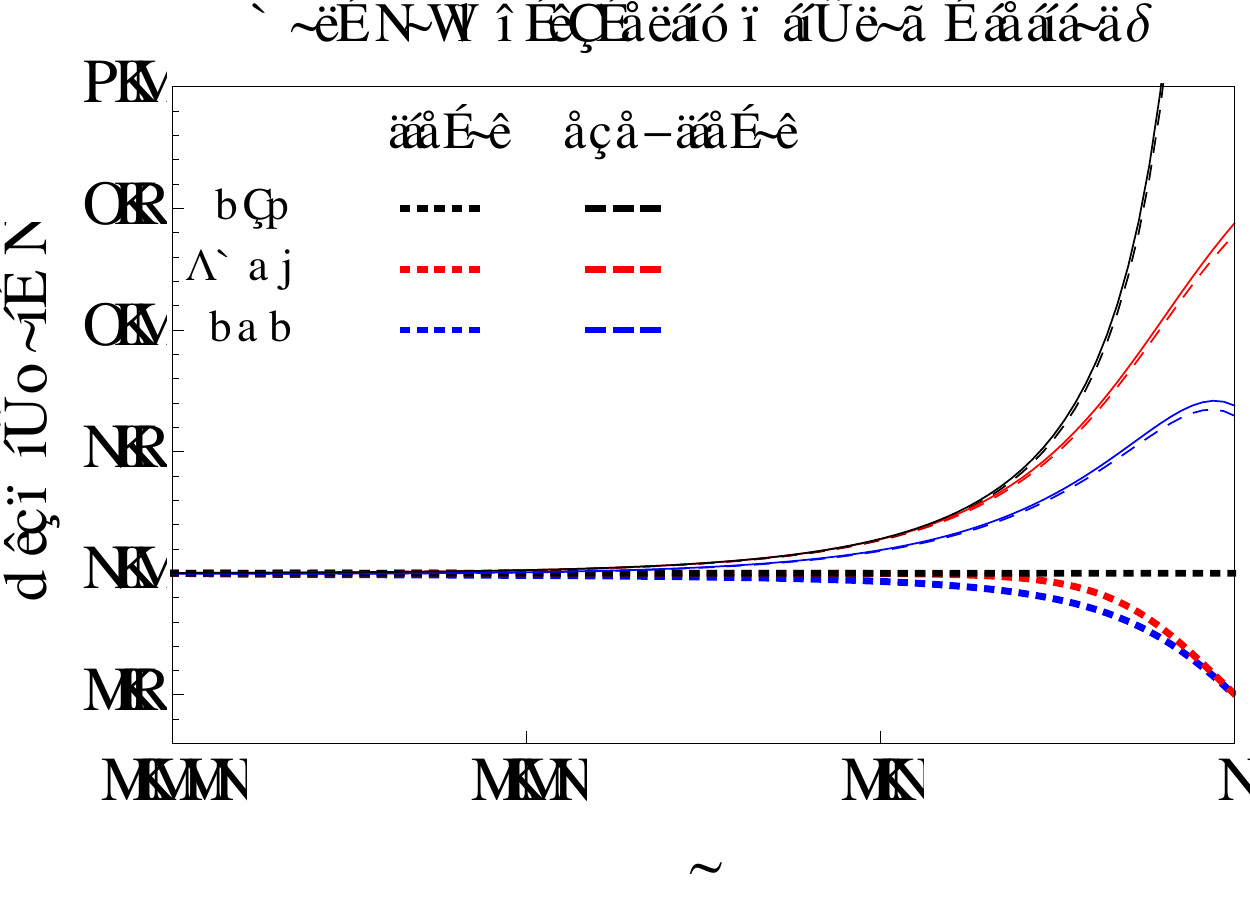} \hspace{1cm}
\includegraphics[width=7cm]{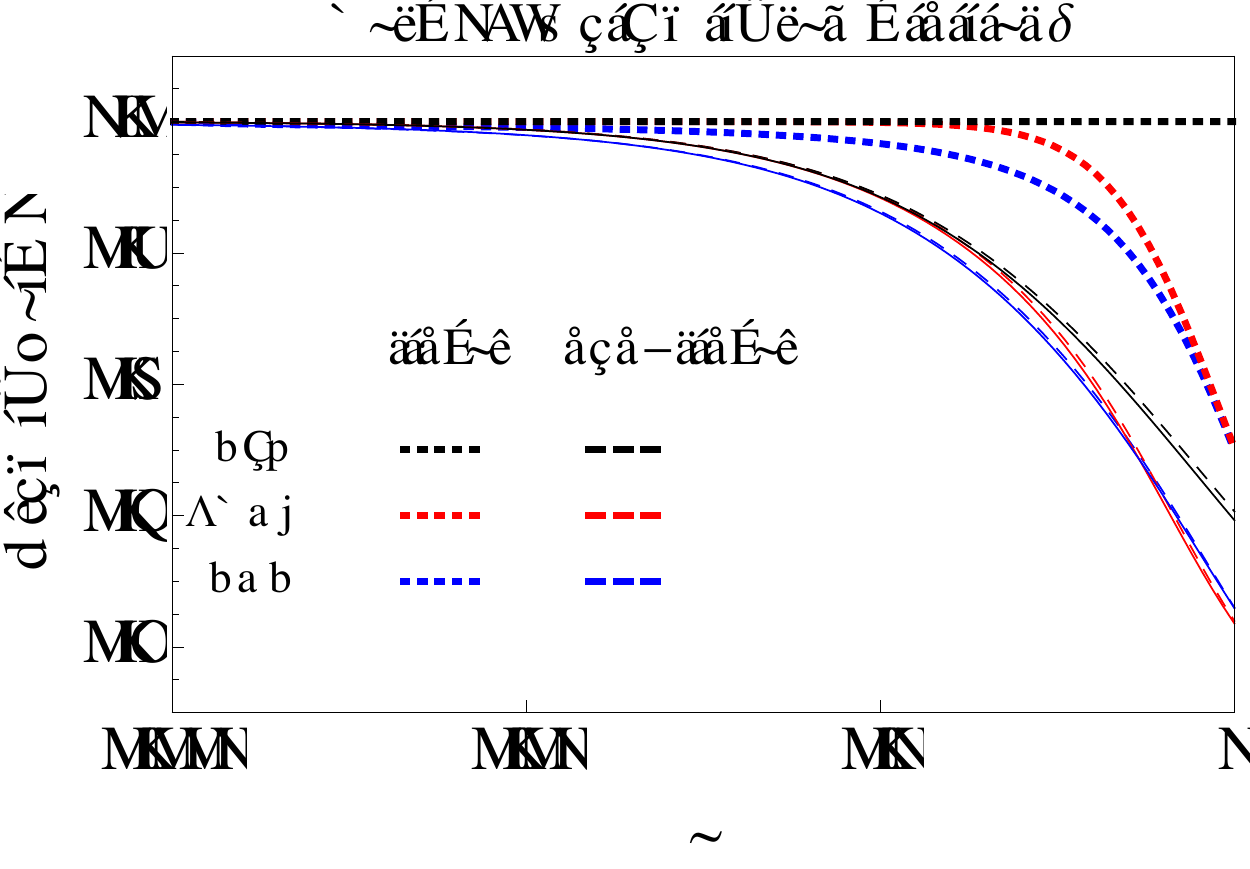}\\
\includegraphics[width=7cm]{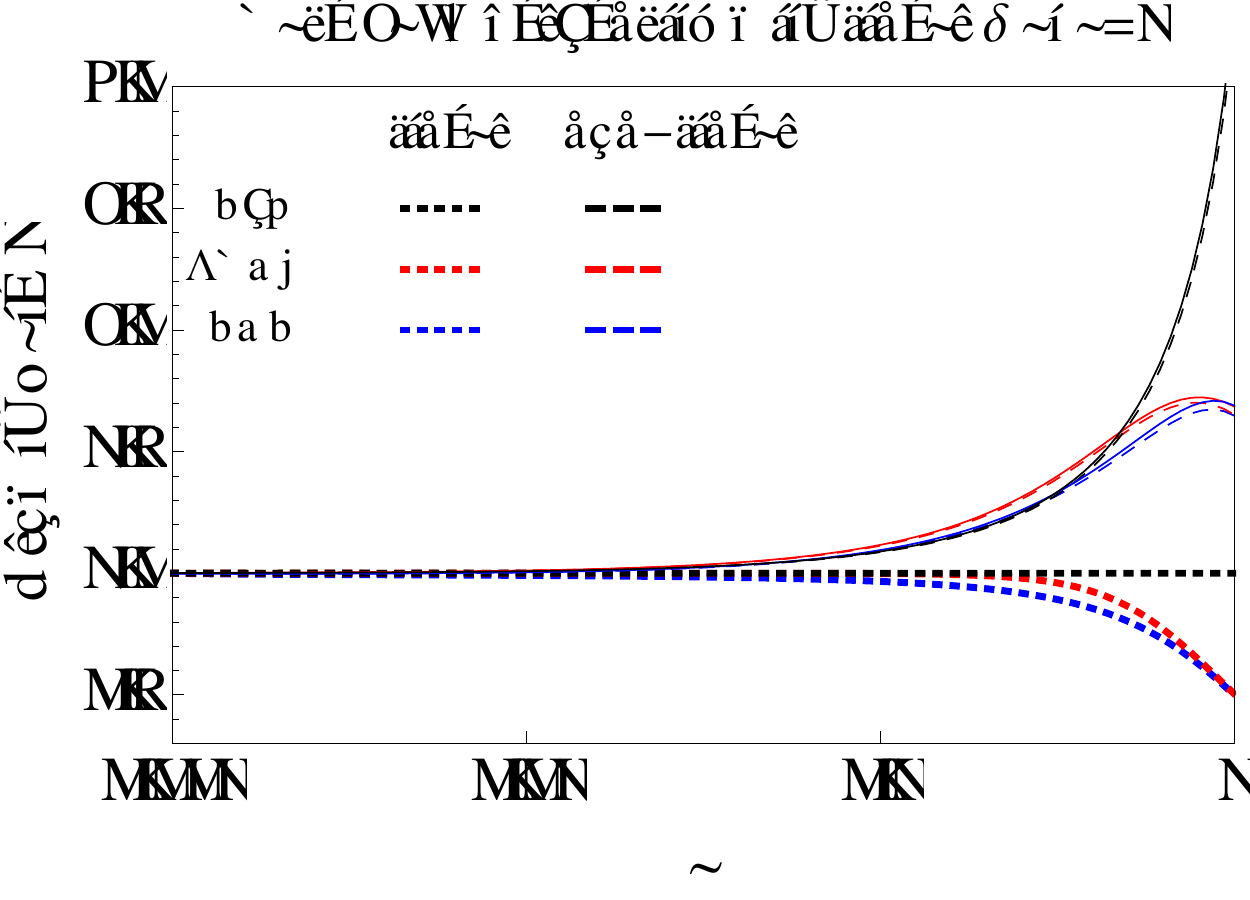} \hspace{1cm}
\includegraphics[width=7cm]{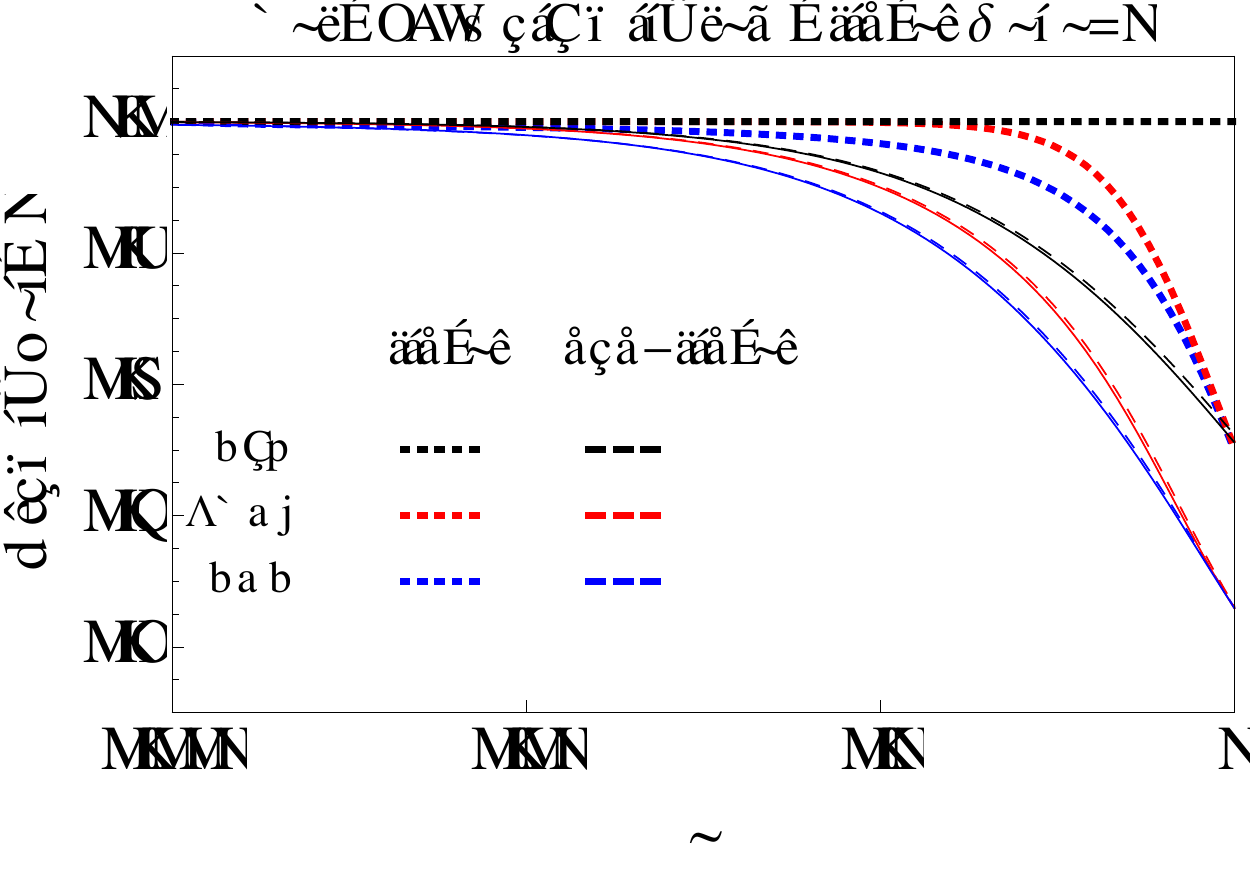}
\caption{ The linear and non-linear growth rates for the four cases discussed in the text. The solid lines indicate the fit given by \eqnref{growthfit}. The EdS model is plotted for reference. The linear growth rate varies with cosmology, but is independent of the initial conditions. 
The growth rates for EDE and $\Lambda$CDM match at $a=1$ because the two models have the same $\Omega_m$ today. In overdensities, in case (1a), the difference in growth rates is purely due to the difference in $\Omega_m$ in the three cosmologies. The cosmology dependence between $\Lambda$CDM and EDE is damped in case (2a) because the EDE cosmology has a higher initial amplitude which compensates the effect of a lower $\Omega_m$. Voids, on the other hand, show a different behaviour. A higher $\Omega_m$ means that the linear growth rate is higher and a void goes non-linear earlier. However, unlike overdensities, the growth rate slows down in the non-linear regime ($f \propto \Omega_m^{\gamma_1}(1+\delta)$). Thus, the growth rate is `self-regulatory' in voids. This explains the small differences between the three cosmologies in case (1b). In case (2b), the initial amplitude for the $\Lambda$CDM (and EdS) is less negative than the corresponding value in case (1b). Thus, at initial epochs, the differences in cosmology are more pronounced, but at late epochs, the differences reduce due to the `self-regulation'.  }
\label{fig:nonlinrate}
\end{figure}

\capfigref{fig:nonlinrate} shows the growth rates for the above four cases. We note the following features. The linear growth rate $d \ln \delta/d \ln a$ does not depend on the initial amplitude but depends only on the cosmological parameters. Thus, it remains unchanged in all four panels. On the other hand, the non-linear growth rate depends on the instantaneous $\delta$, which in turn depends on the initial amplitude as well as the cosmological parameters. For overdensities, the difference between EDE and $\Lambda$CDM is more pronounced when the constraint of `same linear $\delta$ today' {\it is not} imposed. This is similar to the effect seen in \figref{fig:solns} for the same reason. 
 
 However, the case for voids is somewhat counter-intuitive when contrasted with overdensities and \figref{fig:solns}. The growth rate for voids does not show stark differences in behaviour when comparing cases (1b) and (2b). This can be understood as follows. As noted earlier, the more underdense the void, the slower it grows. From \eqnref{growthfit}, the void limit ($\delta \rightarrow -1$) gives,  $f_{void} \propto \Omega_m^{\gamma_1} (1+\delta)$. Thus, for the same initial conditions, a higher $\Omega_m$ implies a faster growth which implies a higher amplitude underdensity which in turn reduces the growth rate i.e., the two factors act in opposite directions.  This `self-regulation' does not occur in overdensities: $f_{overdense} \propto \Omega_m^{\gamma_1+\gamma_2} \sqrt{\delta}$, and the two factors act in tandem. Thus, for the cases of `same initial conditions' (cases (1a) and (1b)), the differences in cosmology are apparent for overdensities but not for voids. In case (2b), the initial amplitude for $\Lambda$CDM and EdS is smaller (less negative) than their values in case (1b). The growth rate is therefore initially larger making the effect of cosmology slightly more pronounced at early epochs, but the difference is reduced at later epochs when the rate slows down as the underdensity grows. Overall, we conclude that cosmology dependence of the growth factor in the overdense case is more sensitive to how initial conditions are set as compared to the void case.  

\begin{figure}
\centering
\includegraphics[width=8cm]{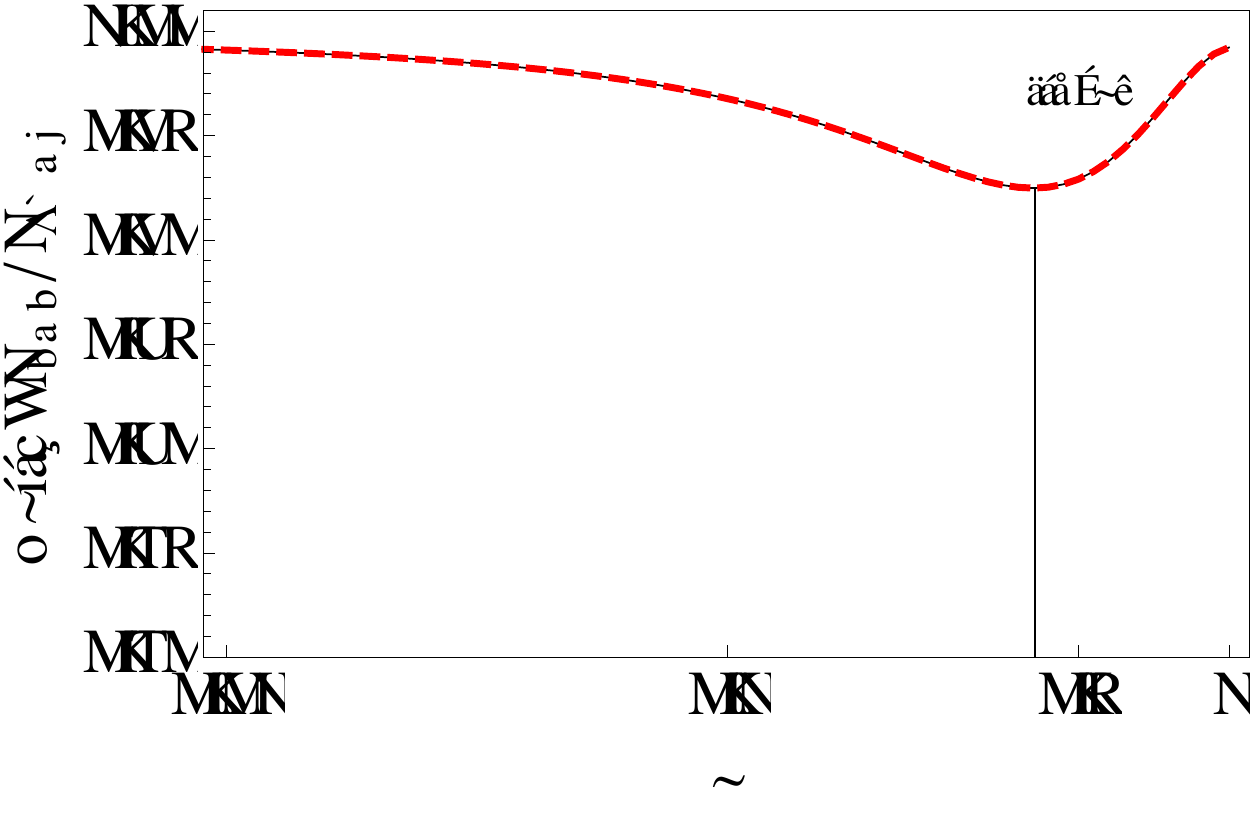}
\caption{Ratio of the linear growth rates. This ratio is independent of the initial conditions since it depends only on the cosmological parameters. The ratio is minimum near $a_{min} \sim 0.4$, which corresponds roughly to the epoch where the ratio of matter density parameters is minimum; the slight offset indicates a weak dependence on the equation of state $w$.   }
\label{fig:linrate}
\end{figure}

The solid lines in \figref{fig:nonlinrate} indicate the fits given by \eqnref{growthfit}, where $\delta$ is evaluated using the numerical solution. We found that over the full range of epochs, the maximum relative error between the numerical value and the value given by the fit was less than 4 \%, indicating that the form given by \eqnref{growthfit} also works well for the EDE model considered here. From the fitting functions, it is clear that the dependence on $w$ is very weak, even in the non-linear regime. In the linear regime, it is usually neglected to give $f_{lin} \sim\Omega_m^{0.55}$ for most dark energy cosmologies with standard gravity. We found that, for the EDE cosmology, ignoring the $w$ dependence (set $w=-1$) gave only an error of 0.8\% (0.4\%) in the non-linear (linear) growth rate. This indicates that the difference in growth rates between $\Lambda$CDM and EDE cosmology is primarily from the $\Omega_m$ dependence. \capfigref{fig:linrate} shows the ratios of the linear growth rates in the EDE to the $\Lambda$CDM cosmology. The black vertical line indicates the position where the ratio $\Omega_{m,EDE}/\Omega_{m,\Lambda CDM}$ is minimum i.e., $a\sim 0.4$. The ratio of the linear growth rates also reaches its minimum close to $a\sim 0.4$. 

%
%
\section{One-point PDF of the density field}
\label{sec:PDF}
\subsection{Numerical Runs}
\label{sec:runs}

The numerical data was generated as follows. An ensemble of points was drawn from a Gaussian random field with mean $0$ and standard deviation $\sigma_G$. This corresponds to the linear distribution at $a=1$. The distribution at any other epoch is obtained by scaling each data point by the normalized linear growth factor, ${\tilde D}(a)$, given in appendix \ref{app:eqns}.  The resulting distribution is Gaussian with the linearly scaled width
\beq 
\sigma_L(a) = \sigma_G {\tilde D}(a). 
\label{linearsigma}
\eeq
The realizations at $a=1$ are the same for all three cosmologies, but differ at the initial epoch due to differences in the growth factors. This set-up is similar to cases (2a,b) in the previous section.
We chose the initial epoch at $a = 0.001$. We considered three values for $\sigma_G: 0.5,\,1$ and $2$. For a BBKS power spectrum with $n_s = 1$ and $\sigma_8 =0.9$, this corresponds roughly to 15, 7 and 3  $h^{-1}$ Mpc. Each initial condition of the ensemble was evolved using \eqnref{deltaNLeqn2} and the final density was read out at $a=0.1,\,0.4$ and $1$ corresponding to redshift $z=9,1.5$ and $0$ respectively. This procedure gives an ensemble of final densities, whose PDF is constructed by first imposing a cut-off $\delta \leq 5$ and then binning with 40 bins equispaced on the log-scale. The PDF was defined as the fractional number of points in each bin divided by the length of the bin. Thus, the resulting PDF is normalized by construction. For each value of $\sigma_G$, five realizations each containing $5 \times 10^4$ points were evolved. The final PDF plotted below is the average over these five realizations and error bars correspond to the standard deviation.

The PDF obtained through this procedure assumes that the mass of each sphere is unchanged throughout the evolution and hence gives a Lagrangian estimate of the PDF. The numerical forms given below have been obtained using Eulerian perturbation theory or are based on simulations and provide the Eulerian PDF. The transformation between the Lagrangian to Eulerian PDF is outlined in appendix \ref{app:lagtoeul}. However, it is important to note that we are not comparing our results to data from N-body codes. In view of this, it is possible to treat the fitting forms given below as purely analytic expressions whose parameters are estimated from our data generated using spherical dynamics. One can then also compare both the Lagrangian and Eulerian PDFs to these forms independently. The exception is B94, which is explicitly Eulerian and does not have any free parameters. It needs to be converted appropriately to facilitate comparison.

\subsection{Fitting forms for comparison}
\label{sec:fittingforms}
In this paper, we have considered three fitting forms (1) the distribution based on Eulerian perturbation theory given by \citet{bernardeau_effects_1994}; hereafter referred to as the B94 distribution, (2) the skewed log-normal distribution (SLN) which combines the Edgeworth series and the log-normal form \citep{colombi_skewed_1994} and (3) the generalized normal distribution ($\mathcal{N}_{v2}$) based on numerical simulations  given by \citep{shin_new_2017}. We compare the analytic expressions to the PDF generated from our numerical runs. 

\subsubsection*{\bf Estimates based on Eulerian Perturbation Theory}
\hspace*{0.5 cm} The B94 is a piecewise distribution for density field and given by,
\beq
\label{theoB94}
P(\delta) = \left
 \{\begin{array}{cc}
P_{B94}^{I}(\delta) &-1\leq \delta < 0.3 \\\\
P_{B94}^{II}(\delta) &\delta >0.3,
\end{array}
\right.
\eeq
where
\beq
P_{B94}^{I}(\delta)d\delta=\Big(\frac{7-5(1+\delta)^{\frac{2}{3}}}{4\pi \sigma_{\delta}^2}\Big)^{\frac{1}{2}}(1+\delta)^{-\frac{5}{3}} \, \text{exp}\Big[-\frac{9}{8\sigma_{\delta}^2}\Big(-1+\frac{1}{(1+\delta)^{\frac{2}{3}}}\Big)^2\Big]d\delta
\eeq
\beq
P_{B94}^{II}(\delta)d\delta=f_c\frac{3a_{s\delta} \sigma_{\delta}}{4\sqrt{\pi}}(1+\delta)^{-\frac{5}{2}} \, \text{exp}\Big[\frac{-|y_{s\delta}|\delta+|\phi_{s\delta}|}{\sigma_{\delta}^2}\Big]d\delta,
\eeq
where $\sigma_\delta = \sigma_L$, where $\sigma_L$ is given by \eqnref{linearsigma}, $a_{s\delta}=1.84$, $y_{s\delta}=-0.184$, $\phi_{s\delta}=-0.03$. We have chosen the $n=-3$ values for the parameters $a_s$, $y_s$, $\phi_s$ (B94). The correction factor $f_c$ is 
\begin{equation}
f_c=[1+2(0.8-\sigma_{\delta})\sigma_{\delta}^{-1.3}(1+\delta)^{-0.5}]
\end{equation}
accounts for the fact that the PDF based on perturbation theory does not perform well at high $\delta$. 

The B94 distribution is derived in the Eulerian frame. Converting to its Lagrangian form involves inverting the Lagrangian to Eulerian mapping given by \eqnref{LtoE} in the appendix. This is not straightforward and as a simplification we use the approximation $P_L(\delta)= P_E(\delta)(1+\delta)$ \citep{nadkarni-ghosh_phase_2016}.

\subsubsection*{\bf Skewed log-normal distribution (SLN)}
\hspace{0.5 cm} The skewed log-normal PDF ($P_{SLN}$) proposed by \citet{colombi_skewed_1994} combines the log-normal PDF and the Edgeworth series. Define $\nu \equiv \Phi/\sigma_{\Phi}$, with $\Phi \equiv \ln (1+\delta)-\langle\ln(1+\delta)\rangle$, $\sigma_{\Phi}\equiv\sqrt{\langle\Phi^2\rangle}$, where the brackets denote the average over the $\phi$ values. $P_{SLN}$, which uses the third order Edgeworth expansion, is defined as

\beq
P_{SLN}(\nu)d\nu=\Big[1+\frac{1}{3!}T_3\sigma_{\Phi}H_3(\nu)+\frac{1}{4!}T_4\sigma_{\Phi}^2H_4(\nu)+\frac{10}{6!}T_3^2\sigma_{\Phi}^2H_6(\nu)\Big]\mathcal{N}(\nu),
\eeq
where
\beq
T_3\equiv\frac{\langle\Phi^3\rangle}{\sigma_{\Phi}^4}, 
\eeq
\beq
T_4\equiv\frac{\langle\Phi^4\rangle-3\sigma_{\Phi}^4}{\sigma_{\Phi}^6},
\eeq
\beq
\mathcal{N}(\nu)=\frac{1}{\sqrt{2\pi}}\exp\Big[-\frac{\nu^2}{2}\Big] 
\eeq 
and $H_m(\nu)$ is the Hermite polynomial of degree $m$. This distribution is not always positive definite, i.e., it can be negative valued hence is not a real PDF \citep{colombi_skewed_1994}. But, N-body simulations show that this approximation successfully describes the late-time evolution of the dark matter field \citep{ueda_counts--cells_1996,szapudi_recovering_2004}. 


\subsubsection*{\bf Generalised normal distribution of version 2 ($\mathcal{N}_{v2}$)}
This distribution has been recently proposed by \cite{shin_new_2017} based on numerical simulations of five different cosmological models. The $\mathcal{N}_{v2}$ distribution consists of three parameters namely, $\alpha$ (scale), $\kappa$ (shape) and $\xi$ (location). $\alpha$ is related to the width of the distribution, $\kappa$ quantifies skewness and $\xi$ gives the location of the peak. In terms of these parameters, the $\mathcal{N}_{v2}$ is defined as,
\begin{equation}
\mathcal{N}_{v2}(\delta)=\frac{\mathcal{N}(y)}{\alpha-\kappa(\delta-\xi)}, 
\end{equation}
where $N(y)$ is the usual Gaussian distribution with mean 0 and variance 1. Here $y$ is defined as,
\beq
y = \left
 \{\begin{array}{cc}
-\frac{1}{\kappa}\ln \Big[1-\frac{\kappa(\delta-\xi)}{\alpha}\Big], & \text{ if $ \kappa\neq 0$} \\\\
\frac{\delta-\xi}{\alpha}, & \text{ if $\kappa = 0$},
\end{array}
\right.
\eeq
A positive (or negative) value of $\kappa$ yields a left-skewed (or right-skewed) distribution bounded to the right (or left). The mean ($\mu$), median (${\tilde \delta}$) and skewness ($\gamma_1$) of the distribution are related to $\alpha$, $\kappa$ and $\xi$ as 
\bea
\label{mean} 
\mu&=&\langle \delta\rangle =\xi-\frac{\alpha}{\kappa}\left(e^{\kappa^2/2}-1\right) \\
\label{median}
{\tilde \delta}&=&\xi\\
\label{skewness}
\gamma_1&=&\frac{\langle(\delta-\mu)^3\rangle}{(\langle(\delta-\mu)^2\rangle)^{\frac{3}{2}}}=\frac{3e^{\kappa^2}-e^{3\kappa^2}+2}{\left(e^{\kappa^2}-1\right)^{\frac{3}{2}}}\rm{sign}(\kappa).
\eea
Here the `$\langle \cdot \rangle$' denotes average over the domain of $\delta$. $\mathcal{N}_{v2}$ approaches the usual Gaussian normal distribution in the limit that $\kappa$ tends to zero. In this case, $\mu = \xi$ and $\gamma_1=0$. 
There are two ways to find the parameters $\alpha$, $\kappa$ and $\xi$. One way is to compute $\mu, {\tilde x}$ and $\gamma_1$ from the data and solve \eqnrefs{mean}, \eqnrefbare{median} and \eqnrefbare{skewness} for the three parameters. The other way is to find the best fit values by optimizing the fit to the data. We implemented the latter method using Python's SciPy inbuilt function. This method requires an initial guess for $\alpha$, $\kappa$ and $\xi$. We found that the errors were minimum when the initial guess values were computed analytically using the first method. Table \ref{table:bestfit} in appendix \ref{app:bestfitparams} shows the values of parameters for the best fit. 

\subsection{Results and discussion}
\begin{figure}
\includegraphics[width=\linewidth]{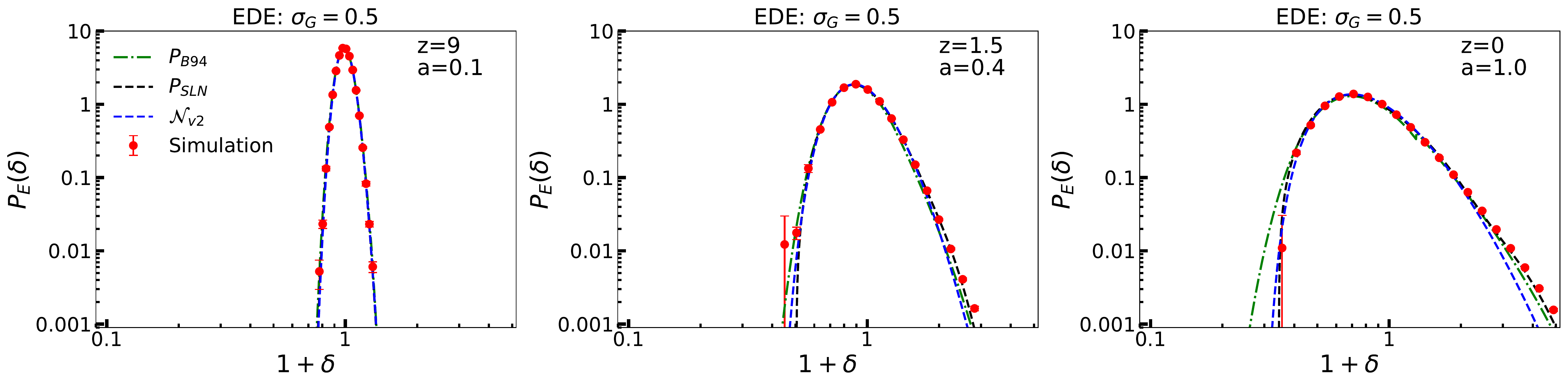}\\
\includegraphics[width=\linewidth]{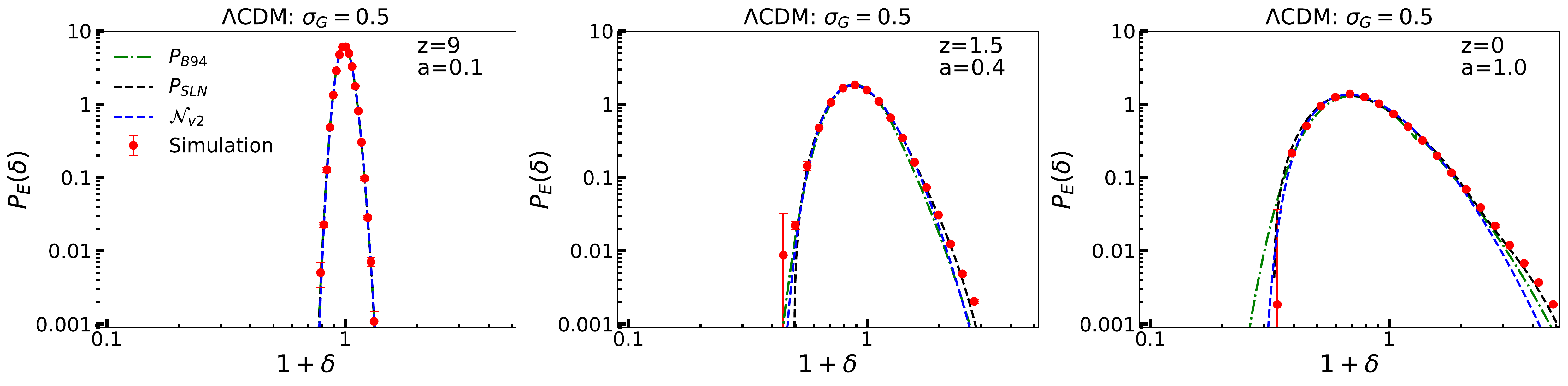}
\includegraphics[width=\linewidth]{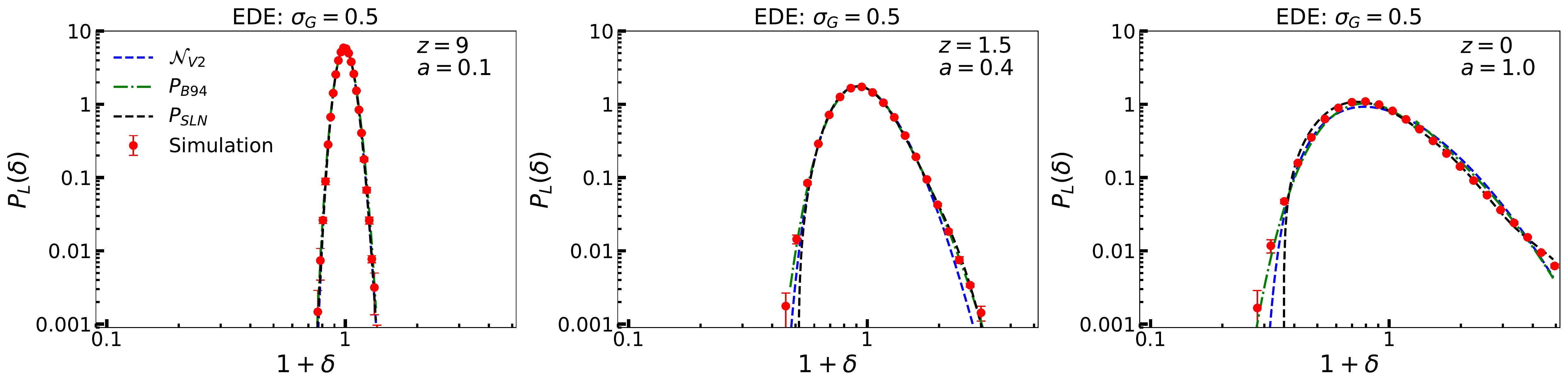}\\
\includegraphics[width=\linewidth]{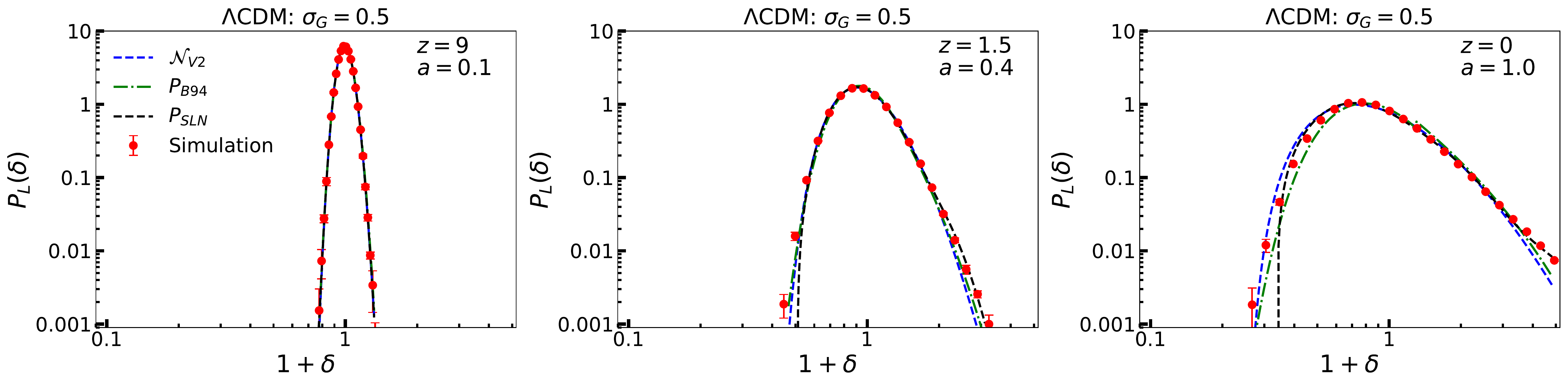}
 \caption{Variation of the one-point Lagrangian space PDF with epoch. $P(\delta)$ vs. $(1+\delta)$ for $\sigma_G=0.5$ for the EDE and $\Lambda$CDM models at $a=0.1$, $0.4$, $1$ ($z=9$, $1.5$, $0$). Simulated PDFs (filled red circles) are compared with the predicted PDFs based on theory and N-body simulations. At $a=0.1$, the Gaussian shape is retained, but non-Gaussian features appear at later epochs.  }
 \label{withredshift}
\end{figure}

\begin{figure}
\includegraphics[width=\linewidth]{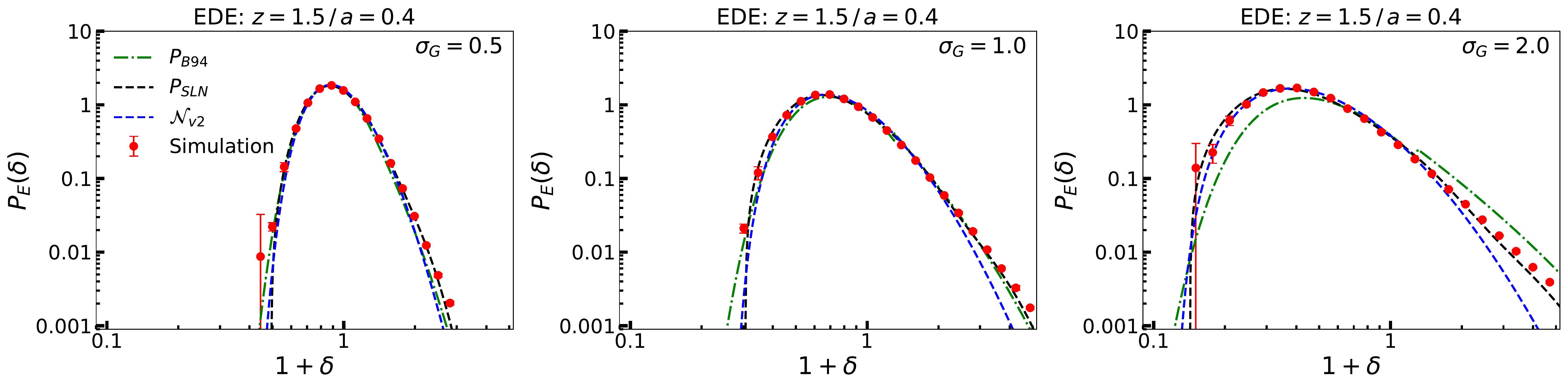}\\
\includegraphics[width=\linewidth]{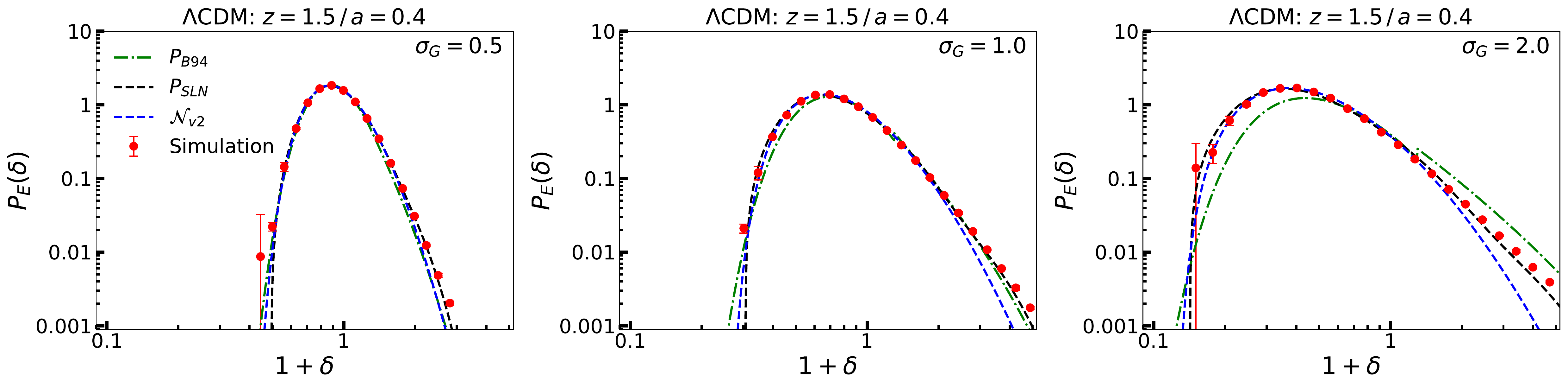}\\
\includegraphics[width=\linewidth]{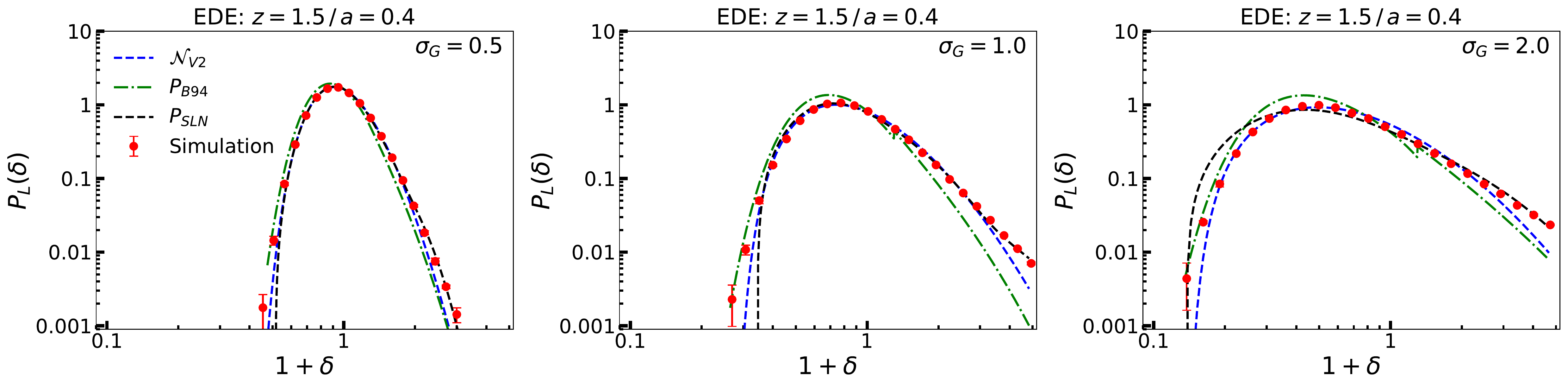}\\
\includegraphics[width=\linewidth]{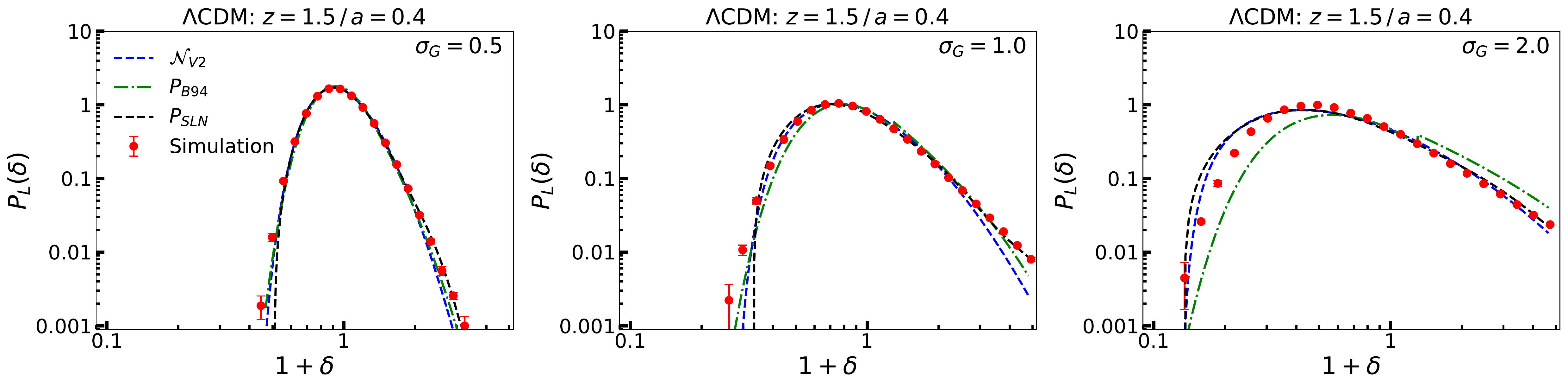}
 \caption{Dependence of one-point Lagrangian space PDF on $\sigma_G$. $P(\delta)$ vs. $(1+\delta)$ at $a=0.4$ ($z=1.5$) for the EDE and $\Lambda$CDM models for three values of $\sigma_G =0.5, 1, 2$. Simulated PDFs (filled red circles) are compared with the predicted PDFs based on theory and N-body simulations. Higher $\sigma_G$ distribution is more non-Gaussian at the same epoch. }
\label{withscale}
\end{figure}

We compute the numerical Lagrangian ($P_L$) and Eulerian ($P_E$) one-point PDFs as described in \S \ref{sec:runs} and compare with the fitting forms described above. The shape of the PDF evolves with time and depends upon $\sigma_G$. \capfigref{withredshift} shows the variation of the PDFs with epoch for $\sigma_G = 0.5$ and \figref{withscale} shows the dependence on $\sigma_G$ for a fixed epoch ($a=0.4$). The top two panels of each figure show $P_E$ and the bottom two show $P_L$.
From \figref{withredshift}, it is clear that at $a=0.1$, the evolution is still approximately linear and the width is given by $\sigma_L$ defined in \eqnref{linearsigma}. Non-Gaussianity sets in at later epochs. The width of the PDF increases with increasing epoch for a given $\sigma_G$. At $a=1$, in the non-linear regime, the SLN distribution provides the best fit to $P_E$ in both the overdense and void regions. SLN also fits $P_L$ in the overdense region, but does not match $P_L$ in voids. 
From \figref{withscale}, it is clear that the SLN form continues to be a good fit to $P_E$ for higher values of $\sigma_G$ in both overdense and void regions. As was the case in \figref{withredshift}, SLN fits $P_L$ in overdensities also for higher values of $\sigma_G$ but fails in voids. The B94 form fits $P_L$ for small values of $\sigma_G$, over the entire domain, but fails for larger values of $\sigma_G$. This breakdown of B94 is expected since Eulerian PT and possibly the approximation for $P_E$ to $P_L$ conversion both break down in the non-linear regime. Overall, we conclude that the SLN form fits the Eulerian PDF best over the entire range of mass scales and redshifts. 
The Lagrangian PDF matches B94 for small values of $\sigma_G$ over the entire redshift range, but for larger values of $\sigma_G$ no particular form fits the Lagrangian PDF well.

The SC is a local model, i.e., each perturbation is assumed to evolve independently. It does not account for tidal forces which cause shear, rotation and accretion. It also ignores non-linear mode coupling, which is a non-local effect. Yet, it recovers the shape of the PDF given by perturbative methods and simulations reasonably well, over a wide range of epochs, scales and cosmologies. The SC model has been successful in other contexts as well. The non-linear relation between the density and velocity field obtained by simulations is a scatter plot, but the `mean' relation is well described by the SC model \citep{bernardeau_non-linearity_1999,kudlicki_reconstructing_2000,bilicki_velocity-density_2008}. 
It has also been successfully used to treat the shell crossing regime in Lagrangian perturbation theory \cite{kitaura_cosmological_2013}. 
Some of this success can be attributed to a partial cancellation between various effects that have been neglected. One illustration of this feature is to consider the effect of rotation and shear. Both terms appear in the Raychaudhuri equation, which dictates the evolution of $\Theta$ and consequently the full non-linear evolution of $\delta$. But they appear with opposite signs (\citealt{peebles_large-scale_1980}). Spherical symmetry ignores both terms, giving rise to a partial cancellation. For example, in modelling the non-linear DVDR, the result of simulations agree better with spherical collapse rather than ellipsoidal collapse \citep{nadkarni-ghosh_phase_2016}. This is because ellipsoidal dynamics accounts for only for the shear and does not benefit from the partial cancellation. These arguments are akin to reasons `why the Press- Schecter formalism works so well' (\citealt{monaco_dynamics_1999}).


\section{The PDF: $\Lambda$CDM vs EDE }
\label{sec:comp}
Many past investigations have looked at the effect of an early dark energy component on the non-linear growth of structure, particularly for the type of EDE model considered here. \cite{bartelmann_non-linear_2006} considered the SC model and computed the critical density for collapse $\delta_c$, an important ingredient in excursion set based estimates of the mass function. They found that $\delta_c$ is lower for EDE models and hence for the same observed structure today, the structure must grow earlier in EDE models. They also noticed that, while there were more clusters at an earlier epoch in EDE than $\Lambda$CDM, the growth rate of structure from redshift 1 to 0 was slower in EDE than $\Lambda$CDM. \cite{grossi_impact_2008} performed numerical simulations to test the conclusions based on semi-analytic estimates. They found that no modification was required to the standard mass function formalism and the sensitivity of the halo mass function to EDE was much less than that predicted based on analytic estimates. Similar conclusions were reached by \cite{francis_can_2008}.  \cite{fontanot_semi-analytic_2012} coupled dark matter simulations with semi-analytic models of gas physics to model galaxy formation in EDE scenarios and again found that galaxy properties could only weakly distinguish EDE from $\Lambda$CDM. More recently, \cite{shi_can_2016} performed more consistent simulations, which accounted for the fact that the EDE parameters that best fit the Planck data are different from those for $\Lambda$CDM. They found that the cleanest signature of dark energy was on the shape of the power spectrum and this difference was elucidated only by adopting consistent linear theory initial conditions. In this paper, we examine the possibility of exploiting the PDF to discriminate between the EDE and $\Lambda$CDM cosmologies. We consider two measures based on the PDF: the difference of the PDFs, particularly  in the bulk and the width of the PDF.

\subsection{Difference of the PDFs}

\begin{figure}
\begin{center}
\includegraphics[width=5.5cm]{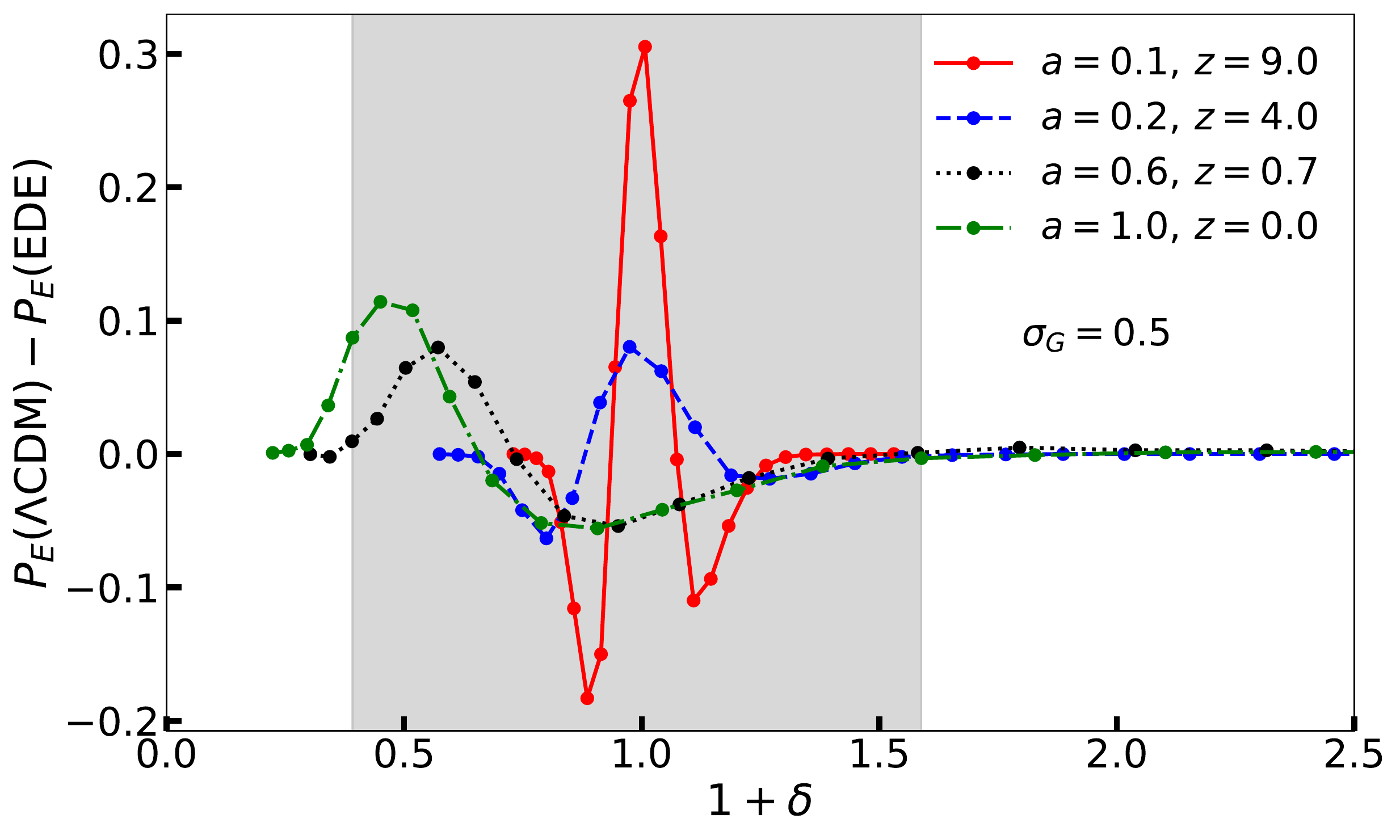}
\includegraphics[width=5.5cm]{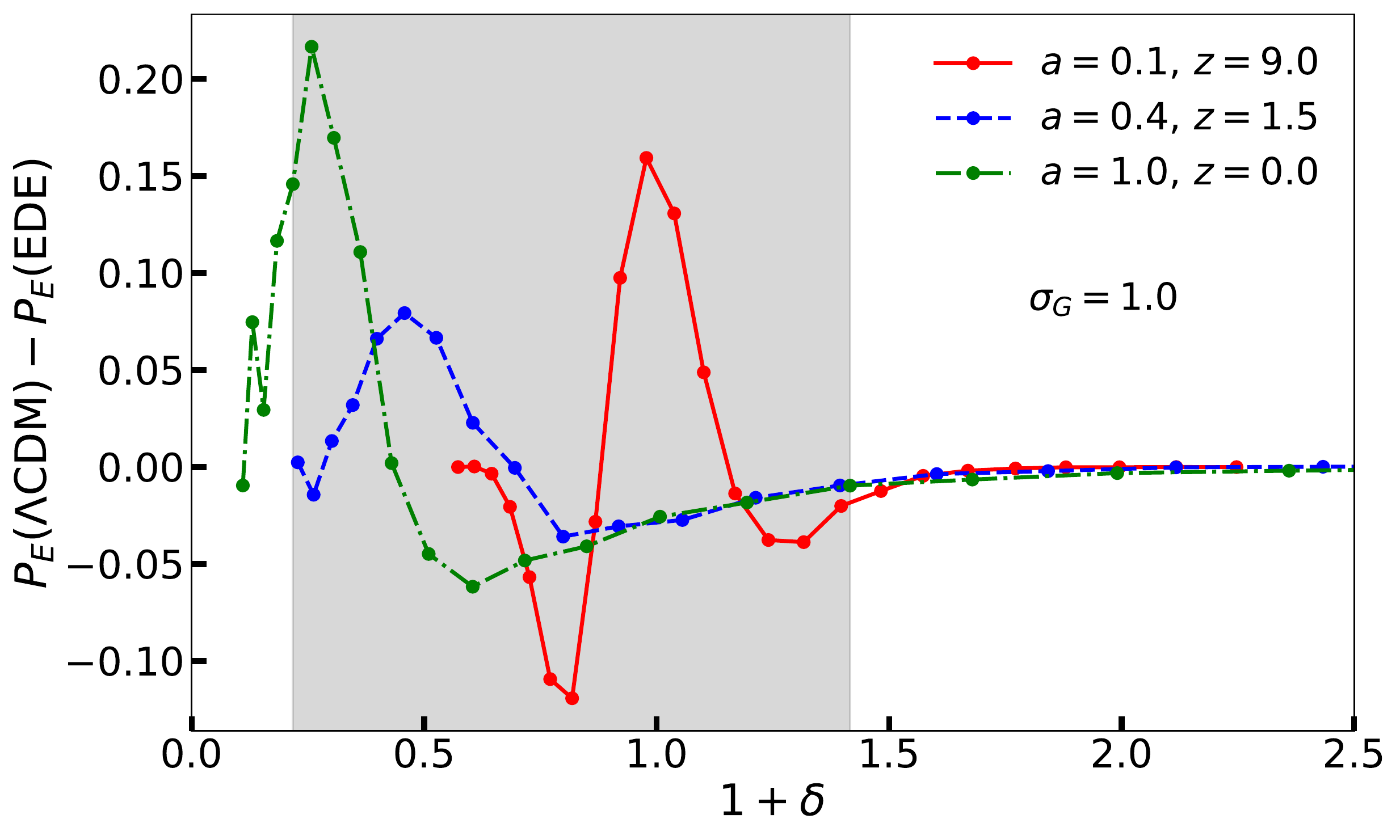}
\includegraphics[width=5.5cm]{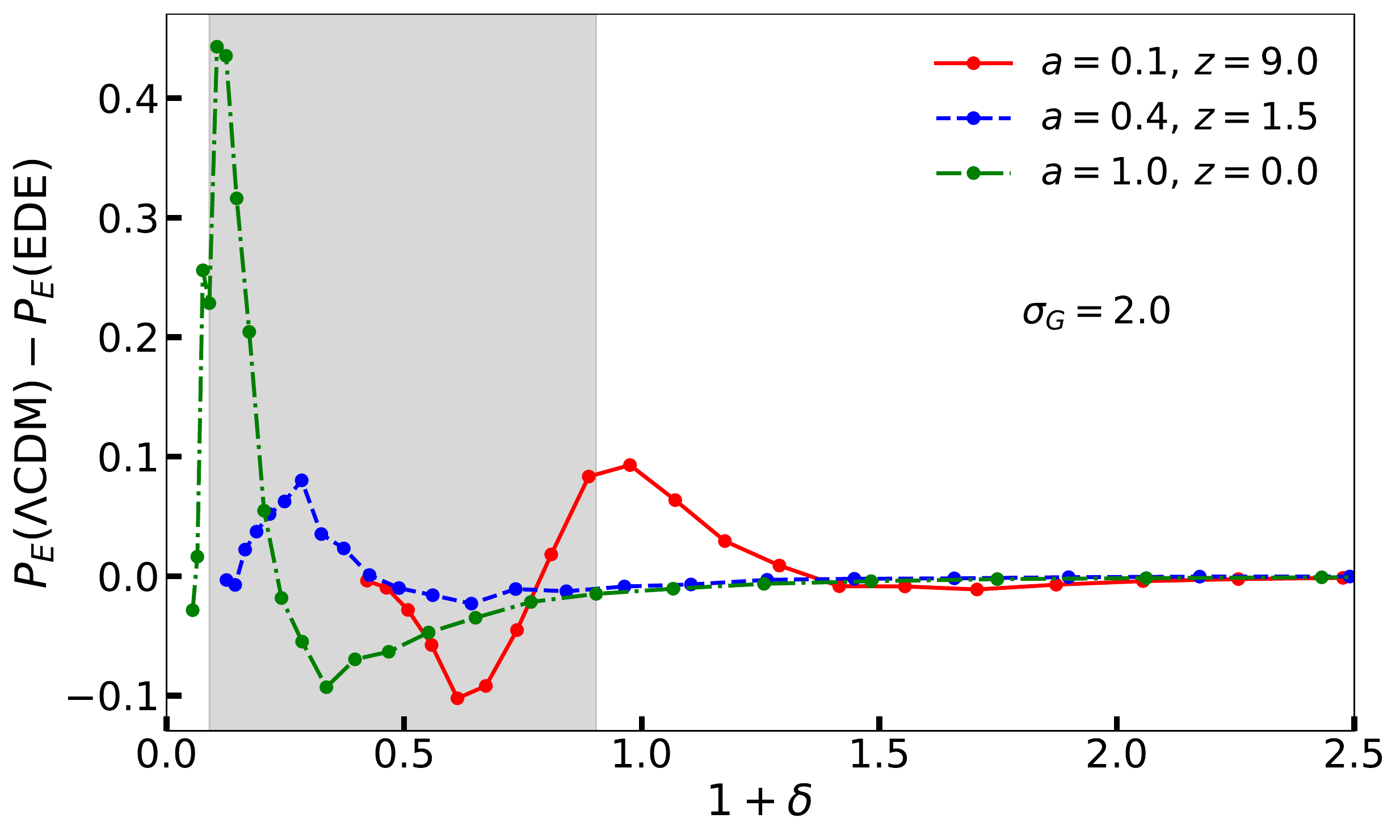}
\caption{Differences between the Eulerian PDF of $\Lambda$CDM and EDE. The grey area denotes the `bulk' region, which is most sensitive to the differences as compared to the tails. The differences near $\delta \sim 0$ are positive at early epochs and reverse sign as evolution progresses. The peak magnitude is higher as non-linear evolution proceeds, but non-monotonic due to the constraint that both cosmologies have the same linear realization at $a=1$. }
\label{differences}
\end{center}
\end{figure}

\capfigref{differences} shows the difference of the Eulerian PDF between the $\Lambda$CDM and EDE cosmology for $\sigma_G = 0.5, 1, 2$. The grey area shows the bulk which is defined as the region around maximum probability which includes 87\% of the area under the curve at $a=1$\footnote{The bulk was defined with respect to the $\Lambda$CDM PDF. We found that data points in the void region which were 10\% less than the peak had huge error bars. Therefore, in defining the bulk, we cut-off the data in the void region which lies below 10\% of the peak PDF. }. We use this definition for all values of $\sigma_G$. It is seen that, at an early redshift, for small values of $\delta$, the difference is positive i.e., the PDF for $\Lambda$CDM is greater than that of EDE and the sign changes with increasing epoch. This can be understood as follows. The initial conditions were chosen such that the realization was the same for the two cosmologies at $a=1$; the values at the starting epoch ($a=0.001$), were chosen by multiplying by the appropriate linear growth factor. Hence, for any point in the realization, the initial value for $\Lambda$CDM is smaller than the corresponding value for EDE. Thus, given the same PDF at $a=1$, the distribution for $\Lambda$CDM at $a=0.001$ will have more points in the $\delta$ bin near zero and hence the difference is positive at the initial time and early epochs. The growth rate is higher for $\Lambda$CDM than EDE and as evolution proceeds, the difference in PDFs for small values of $\delta$ reduces and eventually becomes negative (fewer points in a small $\delta$ bin for $\Lambda$CDM). 

At early epochs, the differences peak (and dip) near $\delta \sim 0$, but the peak moves left as non-linear evolution proceeds. This means the differences between cosmologies are larger in the voids than in the overdense regions. This observation is somewhat counter-intuitive when compared to the discussions in \S \ref{sec:growthrate}, where we conclude that in the non-linear regime, the growth factor in voids is only mildly sensitive to cosmology. Furthermore, for the same fixed linear density contrast today, there was no stark difference between the void and overdense cases. 
This paradox can be resolved by understanding that, although the `self-regulatory' behaviour holds for an individual data point, the cut-off $\delta \geq -1$ means that when an initially symmetric distribution evolves to the non-linear regime, the distribution is no longer symmetric around the origin. The bins in the void region have more data points than those in overdense regions (the overdense points are more spread out) and hence small differences get amplified more in voids. As the evolution approaches $a \sim 1$, the amplitude of the peak gets higher.  At a given red-shift, the peak magnitude is higher for a higher $\sigma_G$ (smaller scale) since there are more points which are non-linear where the differences between EDE and $\Lambda$CDM are more pronounced. 

Recently, \cite{shin_new_2017} also examined the differences of the matter density PDF for four different cosmologies (see their figure 5) and we observe similar magnitudes for the differences at $a=1$. The differences are observed to be larger in voids and the trends with decreasing scale (increasing $\sigma_G$) are also similar.  However, in their case, the magnitudes increase with redshift monotonically, which differs from the behaviour observed here. This is because their method to set up initial conditions differs from ours. The use of the bulk of the PDF to constrain cosmological parameters has also been recently advocated by \citet{friedrich_primordial_2019} in the context of constraining primordial non-Gaussianity. Our definition of the bulk is motivated by this paper and matches it  for $R = 15 h^{-1}$ Mpc. Since we are not using our method to constrain real parameters, we do not rigorously examine the effects of this choice.

\subsection{Width of the PDF}
\begin{figure}
\begin{center}
\includegraphics[height=4cm]{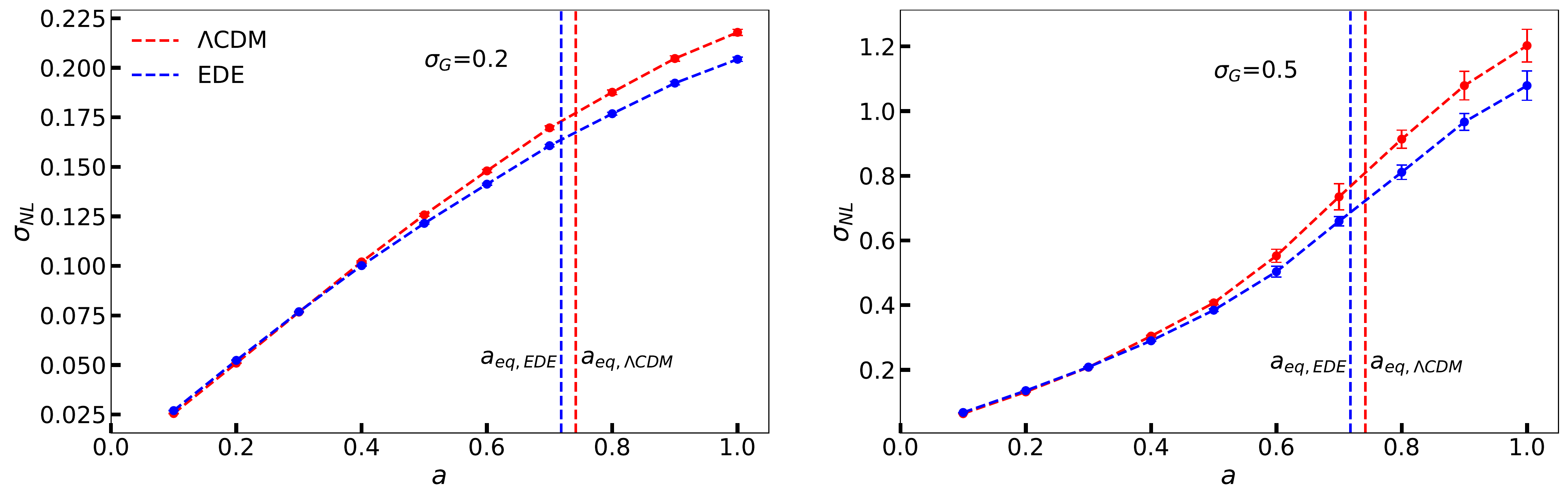}\\
\includegraphics[height=4.2cm]{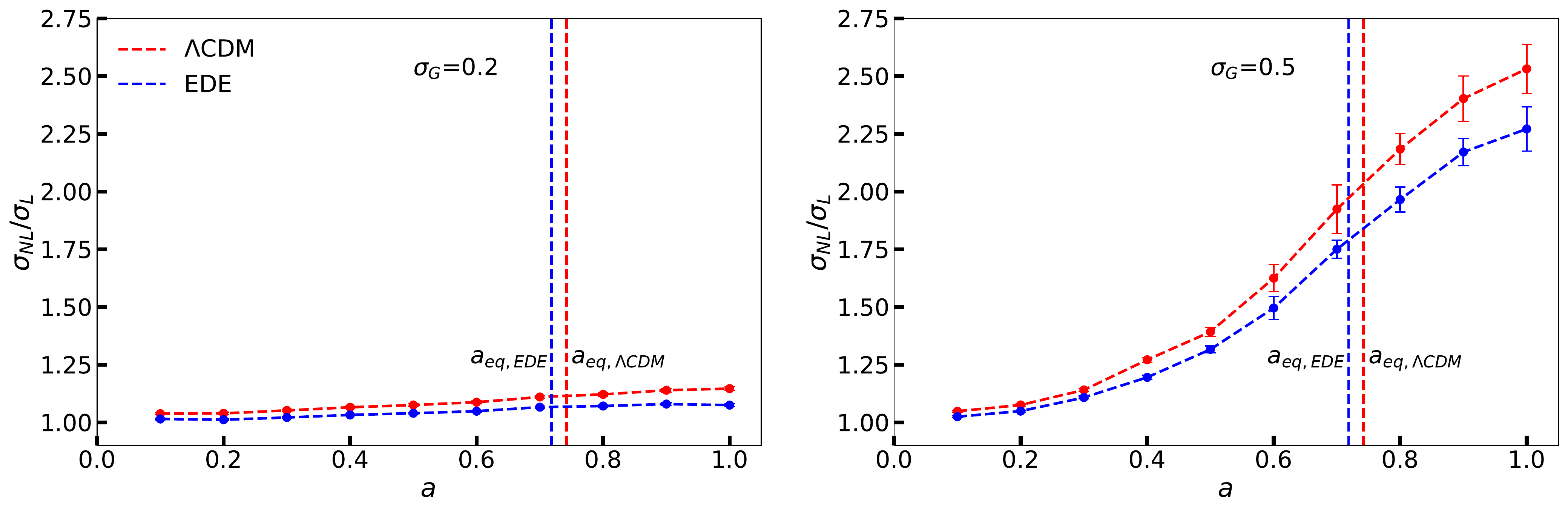}
\caption{Width of the PDF (top panel) and scaled width (bottom panel) as a function of time. At very early epochs, the growth is linear and the scaled width is close to 1. For $\sigma_G =0.2$, the evolution stays close to linear with the $\sigma_{NL}$ deviating from the linear value by only 10\%. For $\sigma_G = 0.5$,  the growth is faster at early epochs but slows down the dark energy terms starts to dominate.}
\label{sigmaNL_fine}
\end{center}
\end{figure}

The width of the PDF is the standard deviation defined as 
\beq
\sigma_{NL}=\sqrt{\langle(\delta-\langle\delta\rangle)^2\rangle}. 
\eeq
The cut-off in the data was imposed at $\delta \leq 20$. In order to highlight the non-linear evolution, we also compute the scaled standard deviation: $\sigma_{NL}$ divided by $\sigma_L$, where $\sigma_L$ is defined in \eqnref{linearsigma}.  \capfigref{sigmaNL_fine} shows $\sigma_{NL}$ (top) and its scaled value (bottom) for two values of $\sigma_G =0.2$ and $\sigma_G =0.5$ (the results for $\sigma_G=1,2$ are subject to the $\delta$ cut-off and are shown in appendix \ref{app:width}). We evaluate the width using the Lagrangian PDF at ten epochs between $a=0.1$ and $a=1$ (the Eulerian PDF showed similar trends). The top panel shows that at early epochs, $\sigma_{NL}$ grows linearly with $a$, for both scales, as expected, since the universe is matter dominated. For $\sigma_G = 0.2$, the fluctuations stay mostly linear, since $\sigma_{NL}$ deviates only about 10\% from its linear value even at late epochs. For $\sigma_G=0.5$, the fluctuations become non-linear at late epochs. The growth rate (slope) increases from its linear value from $a\sim 0.4$ to $a\sim 0.8$, as is expected for non-linear evolution and it slows down after $a\sim 0.8$ due to the presence of the dark energy terms. For both scales considered here, the EDE and $\Lambda$CDM cases have the same qualitative trends. The fractional change between the two cases ($\sigma_{NL,{\rm EDE}}/\sigma_{NL, \Lambda CDM}-1$) at $a=1$ is about 6\% and 10\% for the $\sigma_G =0.2$ and $\sigma_G=0.5$ respectively.

Constraining cosmological parameters is not always straightforward. In general, more than one cosmological parameters can have the same effect on a given observable. These degeneracies imply that parameters cannot always be constrained uniquely. For example, at a single red-shift the effect of EDE can also be mimicked by a lower $\Omega_{m,0}$. Other parameters that can be degenerate with the EDE equation of state parameters are the tilt of the power spectrum $n_s$ or its normalization $\sigma_8$. One interesting way to break this degeneracy is to use phase space analysis. It was shown in N13, the simulated density and velocity profiles of galaxies were degenerate with respect to $\sigma_8$ and $w$. However, plotting the same data points on a $\delta-\Theta$ phase space plot, could break this degeneracy since the late time non-linear relation between $\delta-\Theta$ depends only on the parameters that dictate evolution: i.e., the $\Omega$s and $w$. It is independent of the primordial parameters such as $n_s$ and $\sigma_8$ because the relation is an attracting set of the dynamics. Other degeneracies may arise due to new physics - the presence of massive neutrinos or modifications to gravity and this may complicate parameter estimation. However, investigations of the PDF that include neutrinos suggest that the signature of clustering neutrinos on the PDF is not generate with changes in $\sigma_8$ \citep{uhlemann_fisher_2020}. In the current framework, the initial conditions are characterized only by the initial scale ($\sigma_G$) and it has no free parameters other than the fractional densities and equation of state. Thus, 
 an in-depth analysis of these degeneracies in the context of EDE cosmology is beyond the scope of this work.

\section{Conclusion}
\label{sec:conclusion}

In this paper, we computed the one-point PDF of the dark matter density field using spherical collapse as a proxy for the non-linear dynamics.  We focussed primarily on comparing the standard $\Lambda$CDM model with a three-parameter Early Dark Energy (EDE) model \citep{wetterich_phenomenological_2004,doran_early_2006}. The initial distribution was assumed to be Gaussian characterized by its width $\sigma_G$. The non-linear dynamics of any point in the initial realization is governed by \eqnref{deltaNLeqn2} and the PDF at later epochs is numerically computed from knowing the individual densities at the desired final time.  Since the mass of the sphere is assumed to be a constant of motion, the estimated distribution corresponds to the Lagrangian PDF and it has to be appropriately transformed to the Eulerian PDF (appendix \ref{app:lagtoeul}). This procedure is purely local; it does not account for any non-local interactions between spheres. 
There were three main aims: (1) to test the efficacy of this method by comparing with existing closed forms available in the literature,  (2) to compare the PDFs for EDE and $\Lambda$CDM models and (3) to examine if the density-velocity divergence relation (DVDR) derived by Nadkarni-Ghosh in N13 for constant $w$ cosmologies can also be applied to the EDE model with time-varying $w$. We considered three values of $\sigma_G: 0.5, 1$ and $2$, which roughly correspond to a smoothing scale of $R_f =15, 7$  and 3  $h^{-1}$ Mpc for the BBKS power spectrum and may vary slightly for other choices\footnote{The algorithm does not require a specification of the power spectrum.}.  To compare the PDFs, we considered three fitting functions: the theoretical estimate given by Bernardeau, referred to as B94 \citep{bernardeau_effects_1994}, the skewed lognormal (SLN) form given by \mbox{\cite{colombi_skewed_1994}} and the generalized normal distribution  version 2 ($\mathcal{N}_{v2}$) recently proposed by \mbox{\citealt{shin_new_2017}}, based on numerical simulations.  There are other forms based on large deviation statistics \citep{uhlemann_back_2016} or excursion sets \citep{lam_perturbation_2008} but we do not compare to these forms for want of a closed form fitting function. To compare the EDE and $\Lambda$CDM models, we compute the PDF differences in the bulk and the PDF width as two measures. 
The main conclusions are:

\begin{itemize}
\item The SLN form matches the Eulerian PDF for the entire range of redshifts (from early epochs to today), scales and cosmologies, in both the overdense and void regions. It also provides a good fit to the Lagrangian PDF in the overdense regions but fails in voids. The B94 form (transformed to the Lagrangian frame) matches the Lagrangian PDF for small values of $\sigma_G$ in both overdensities and voids, but for larger values of $\sigma_G$, no form provides a good fit to the entire Lagrangian PDF.

\item The differences between the PDF, were examined in the bulk of the PDF, which contains about 87\% of the probability. The peak differences did not increase monotonically with redshift because the initial conditions were chosen such that the two cosmologies have the same linear distribution at $a=1$. Imposing this constraint damps the effect of the differences in non-linear growth rates in the two models. The peak height also increases with increasing $\sigma_G$. At early epochs, the peak of the difference is at $\delta \sim 0$, but shifts to the void as non-linear evolution progresses. This is because of the inherent asymmetry in the dynamics: $\delta\geq-1$ in voids, but it has no bound in the overdense limit. The temporal evolution of this peak will potentially different for other initial conditions. 

The width of the PDF, defined to be the standard deviation, was computed for $\sigma_G = 0.2 $ and 0.5. The observed differences between the two models were about 10\%. Both measures, differences and the width can be used as potential probes to discriminate between models, however, the differences have a slight advantage. Within this simple model, defining the width for higher values of $\sigma$  is subject to imposed cut-offs. More realistically, theoretical modelling in the tails is complicated by other effects (for example baryonic physics) and  observationally, the tails are harder to measure and are ridden by many uncertainties such as galaxy bias and sample variance. On the other hand, the differences, particularly in the bulk, have information encoded in the peak as well as the amplitude and this additional information can be used to break parameter degeneracies (see for example \citealt{friedrich_primordial_2019}).

\item We computed the growth rate of perturbations in the linear and non-linear regime; the growth as expected is slower in EDE than dark energy. One would naively expect that the difference in the growth rates would be magnified in the non-linear regime as compared to the linear regime. However, the linear growth rate depends only on cosmological parameters whereas the non-linear growth rate depends not only on the cosmological parameters, but also on the instantaneous density contrast and hence in turn is sensitive to initial conditions. Thus, we found that in overdensities, if initial conditions are the same in both models then the growth rates in the non-linear regime are a very sensitive probe of cosmology. However, if the initial conditions are chosen such that the linear contrast  is the same for both models at the current epoch, then the effect of cosmology is damped by this constraint. In voids,  the non-linear growth rate in a void is proportional to $\Omega_m^{0.56}(1+\delta)$. These two factors act in opposite giving rise to a `self-regulatory' behaviour. A higher value of the matter density parameter implies a faster transition to the non-linear regime, however, a higher non-linearity implies a lower value of $(1+\delta)$ bringing down the net growth rate. Thus, the non-linear growth rate in voids is neither significantly altered by the choice of initial conditions, nor is it too sensitive to cosmology. Nevertheless, the small differences between models get amplified in the PDF because the non-linear PDF is skewed with a cut-off at $\delta \geq -1$. For initially symmetric initial conditions (as given by a Gaussian), the differences in the non-linear PDF are higher in voids. The case may potentially be different in the presence of a primordial non-Gaussianity.

\item The form of the non-linear DVDR given by N13 was tested by examining the observed and predicted non-linear growth rates. 
The predicted growth rate was computed by combining the definition given by \eqnref{nonlinrate} with the fitting form of $\Theta$ from N13, given by \eqnref{Thetafit}. 
The error between observed and predicted rate is, less than 4\% confirming the validity of the N13 formula for EDE models. This suggests that the formula is robust even though it was derived based on constant $w$ dark energy models. 
\end{itemize}

Observationally, the one-point PDF is measured either using galaxy catalogs or using weak lensing convergence maps. The survey either measures positions of galaxies or observed ellipticities to compute a shear map. The density map is constructed using the Cloud-In-Cell (CIC) method which involves defining 2D or 3D cells in space, adding up the matter content in each cell and allotting a density to the cell depending upon the cell volume (for e.g., \citealt{bernardeau_large-scale_2002}). This method introduces shot noise because of the finite cell size. This noise is often modelled by a Poisson distribution in the case of galaxies and a Gaussian in the case of the weak lensing convergence field (for e.g., \citealt{clerkin_testing_2017}). In order to compare to data, the theoretically predicted continuous distribution has to be convolved with the noise distribution. In the case of galaxies, one needs to further account for the fact that they are biased tracers of the underlying dark matter distribution. Both these effects may introduce effects that are degenerate with the effect of changing cosmology \citep{szapudi_recovering_2004,manera_local_2011,jee_second-order_2012,uhlemann_question_2018}. 

The use of spherical dynamics to model the density one-point PDF is not new. Many investigations have coupled SC with some non-local techniques like perturbation theory, Edgeworth expansion, excursion sets, large deviation theory, path integral formalism, all of which account for the interaction between modes. However, in this analysis, we do not couple SC with any non-local scheme, which makes this a rather simple technique. Our study is motivated by the future goal of finding the one-point PDF for non-standard dark energy models (for e.g., \citealt{bagla_cosmology_2003,pratap_rajvanshi_nonlinear_2018,sangwan_reconstructing_2018,gupta_thawing_2015,devi_evolution_2011,chakraborty_thawing_2019}) or modified gravity. Complications arise in these situations because the dynamical variables are no longer just density and velocity. In the case of modified gravity, the ratio of the two gravitational potentials also becomes a dynamical variable (Nadkarni-Ghosh {\it et al}, in preparation), whereas in the case of dark energy coupled to dark matter, the dark energy perturbation introduces an additional degree of freedom. Top-hats do not stay as top-hats (e.g., \citealt{dai_consequences_2008,borisov_spherical_2012,kopp_spherical_2013}) and defining the density PDF will involve taking a spatial average of the density for each initial perturbation. These issues can be better understood, with a simple scheme. 
Our work is perhaps most similar in spirit to that of \citet{ohta_evolution_2003} who obtained an analytical expression for the evolution equation of the PDF whereas, we follow the evolution of the PDF numerically. The new features of our analysis are the one-point PDF and the expression for the non-linear growth rate, which is derived from the density velocity divergence relation (eq. \ref{growthfit}). 

The SC model has numerous limitations. It is based on local dynamics. Thus, the two-point correlation functions and higher-order spatial statistics cannot be computed solely within this framework. Effect of the environment which induces tidal forces, rotation and accretion is ignored and so is mode-coupling and galaxy bias (see \citet{fosalba_cosmological_1998-1} for a detailed analysis of ignoring the tidal fields in SC while modelling cumulants). Estimates based on perturbative schemes such as Lagrangian perturbation theory \citep{nadkarni-ghosh_extending_2011,nadkarni-ghosh_modelling_2013} or other approximate methods \cite{monaco_approximate_2016} can account for some of the neglected effects. Nevertheless, the method presented in this paper allows for a quick comparative study of the evolution of the one-point PDF in various cosmologies.  

\section{Acknowledgements}
We would like to thank the anonymous referee for very detailed and valuable comments which improved this work significantly. SN would like to acknowledge the Department of Science and Technology (DST), Govt. of India for the grant no. (SR/WOS-A/PM-21/2018). 

\appendix
\section{The dynamical equations}
\label{app:eqns}
The physical system comprises of a perturbed sphere evolving in a background cosmology consisting of matter and dark energy with an equation of state $w$. The equation of state can be varying or time-dependent. The background evolution is completely described by a scale factor $a(t)$ given by \eqnref{backeqn}. 

Let the origin be at the center of the inner sphere (perturbation) and ${\bf r}$ and ${\bf x}$ denote the physical and comoving coordinate of the outer edge of the sphere. Let ${\bf u} = {\dot {\bf x}}$ and $\delta$ denote the fractional overdensity.  For pure matter cosmologies,  it is possible to write down an analogous Friedmann equation by replacing the scale factor of the background by that of the perturbation with a different matter density parameter. However, the extension to other dark energy models is not obvious.  Therefore, it is preferable to start with the equations of hydrodynamics, generalized to include dark energy.
This set corresponds to the spherically symmetric, sub-horizon limit of the perturbed Einsteins' equations along with the conservation of the stress-energy tensor. 
In the weak-field, sub-horizon limit, for pure matter perturbations, the equations are: 
\bea
\label{cont}{\dot \delta} &=& -(1+\delta) \nabla_x \cdot {\bf u} \\
\label{euler}{\dot {\bf u}} + 2 H  {\bf u}&=& = -\frac{1}{a^2} \nabla_x \psi\\
\label{Nlaw}\nabla_x^2 \psi &= & \frac{3}{2} \ a^2 \Omega_m(a) H^2(a) \delta, 
\eea
where the `dot' refers to the total derivative w.r.t. cosmic time $t$ ($d/dt = \partial/\partial t + {\bf u} \cdot \nabla_x$) and the partial spatial derivatives are with respect to the comoving coordinate ${\bf x}$. Taking divergence of \eqnref{euler}, gives
\beq 
\nabla_x \cdot {\dot {\bf u}} + 2 H \nabla_x \cdot {\bf u} = - \frac{3}{2}H^2 \Omega_m(a) \delta. 
\label{euler2}
\eeq
For the spherically symmetric system considered here, we have 
\beq
\frac{d (\nabla_x \cdot {\bf u} )}{dt} = \nabla_x \cdot {\dot {\bf u} } - \frac{(\nabla_x \cdot  {\bf u})^2}{3}
\eeq
Define  the scaled velocity divergence as 
\beq
\Theta = \frac{\nabla_x \cdot {\bf u}}{H}.
\eeq
 \capeqnrefs{cont} and \eqnrefbare{euler2} give the coupled system 
\bea
{\dot \delta} &=& -H(1+\delta) \Theta \\
{\dot \Theta} &=& -H\left[\frac{3}{2} \Omega_m(a) \delta + \left(2 + \frac{\dot H}{H^2} \right) \Theta +\frac{ \Theta^2 }{3}\right].
\eea
Combining the two first order equations gives the second order system 
\beq
{\ddot \delta} + 2 H {\dot \delta}  -  \frac{4}{3} \frac{{\dot \delta}^2}{(1+\delta)}  - \frac{3}{2} H^2 \Omega_m(a) \delta(1+ \delta) =0.
\eeq
Using $\ln a$ as the time variable instead of $t$ ($d/dt = Hd/d\ln a$), gives the coupled system 
\bea
\delta' &=& -(1+\delta) \Theta\\
\Theta' &=& -\frac{3}{2} \Omega_m(a) \delta - \Theta \left(2 + \frac{H'}{H} \right) - \frac{\Theta^2}{3}
\eea
and the second order version 
\beq
\delta'' + \left(2 + \frac{H'}{H}\right) \delta' - \frac{4}{3} \frac{\delta'^2}{(1+\delta)} - \frac{3}{2} \Omega_m(a) \delta(1+\delta)=0.
\label{app:deltaeq1}
\eeq
For the dark energy cosmologies considered in this paper: 
\beq 
\frac{1}{H} \frac{dH}{d\ln a} = -\frac{3}{2} \Omega_m(a) - \frac{3}{2} (1+ w(a))\Omega_{de}(a).
\eeq 
For a flat universe: $\Omega_m(a) + \Omega_{de}(a) = 1$ at all epochs. So the factor 
\beq
\left(2 + \frac{H'}{H}\right) = \frac{1}{2} \left[1-3w(a) \Omega_{de}(a)\right].
\eeq
This gives 
\beq
\delta'' - \frac{4}{3} \frac{\delta'^2}{(1+\delta)} + \frac{1}{2} \left[1-3w(a) \Omega_{de}(a)\right]\delta' - \frac{3}{2} \Omega_m(a) \delta(1+\delta)=0.
\eeq
This second order non-linear equation for the evolution of $\delta$ has been considered in the past by \cite{fosalba_cosmological_1998} as well as \cite{ohta_evolution_2003}, but applied to pure matter cosmologies. 

\begin{figure}
\centering
\includegraphics[width=16cm]{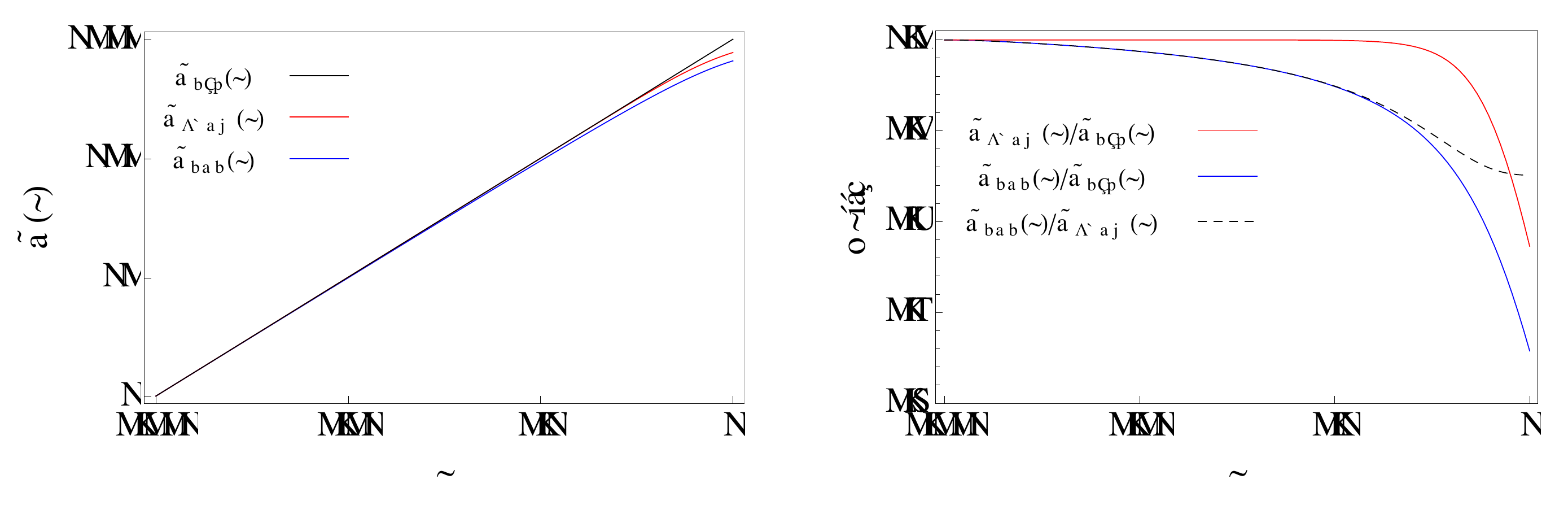}
\caption{Left panel: the growth factor ${\tilde D}$, normalized to unity at the initial time. For the EdS cosmology it is proportional to $a$.Until the dark energy terms become prominent, the other models too behave like the $EdS$ model. Right Panel: Ratios of the growth factors.The EDE and $\Lambda$CDM have a lower growth factor as compared to the EdS case; the EDE cosmology has a slower growth as compared to  $\Lambda$CDM.   }
\label{fig:growthfac}
\end{figure}

In the linear regime, \eqnref{app:deltaeq1} reduces to
\beq
\delta'' + \left(2 + \frac{H'}{H}\right) \delta' - \frac{3}{2} \Omega_m(a) \delta=0
\eeq
The general solution to the linear system is (see equations 7.73, chapter 7 \citealt{dodelson})
\beq 
\delta_{lin}(a)  = C_1 H(a) + C_2 D(a),  
\eeq
where 
\beq 
D(a) \propto H (a) \int_{0}^a \frac{da'}{a'^3 H(a')^3}.
\eeq
Requiring that there be no decaying modes, sets $C_1 =0$. Thus, the linear solution is 
\beq
\delta_{lin}(a) = \delta(a_{i}) \frac{D(a)}{D(a_{i})},
\eeq
where $\delta(a_i)$ and $D(a_i)$ are the initial  values of the density and the growth factor. Define the scaled growth factor 
\beq 
{\tilde D}= \frac{D(a)}{D(a_i) }.
\eeq
\capfigref{fig:growthfac} shows the growth factor and various ratios for the cosmologies discussed in this paper. 
Growth factor is different from the growth rate which is defined as 
\beq 
f = \frac{d \ln \delta}{d \ln a} = \frac{d \ln D(a)}{d \ln a},
\eeq
and is discussed in detail in the main text. 

\section{Comparison between the EDE and CPL models}
\label{app:EDECPLcompare}
 The equation of state for the EDE model considered in this paper is given by \eqnref{eosEDE}. The EDE model contains a low but non-vanishing amount of dark energy at early epoch. It can be characterized by the quantity ${\bar \Omega_{de,sf} }$ defined in \eqnref{omegasf} \citep{doran_structure_2001,doran_impact_2007}
which is the average fraction of dark energy during the structure formation/matter domination era. Using a combination of data sets, including WMAP-3 year data, \cite*{doran_impact_2007} found  ${\bar \Omega_{de,sf} } \leq 0.04$. A randomly chosen model that satisfied these constraints is $\Omega_{de,e} = 8 \times 10^{-4} $ and $w_0 =-0.99$ \cite*{bartelmann_non-linear_2006} . Another popular parameterization of dynamical dark energy models is the CPL parameterization \citep{chevallier_accelerating_2001,linder_exploring_2003}
\begin{equation}
    w(a)=w_0+w_a(1-a) 
    \label{EoSCPL2}
\end{equation}
where $w_0$ and $w_a$ are two free parameters and $w_0$ is to the value of $w$ at present time. Most recent observational constraint from \citet{Plank_collab2018}, which combines the \textit{Planck}, SNe and BAO data  is $w_0=-0.957$ and $w_a=-0.29$. 
\begin{figure}
\centering
\includegraphics[height=4cm]{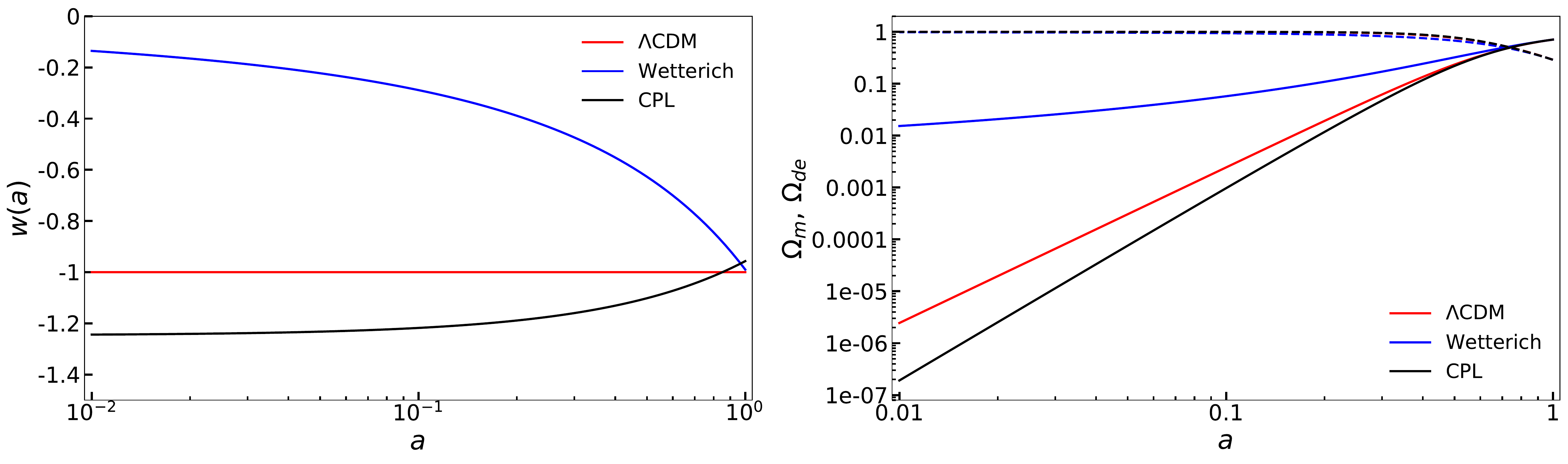}
 \caption{The left panel shows the evolution of the equation of state parameter $w$ and  as a function of scale factor $a$. The right panel shows the evolution of density parameters $\Omega_m (a)$ (dashed line) and $\Omega_{de}(a)$ (solid line) as a function of $a$. The blue and the black lines correspond to the EDE model (with $w_0=-0.99$, $\Omega_{de,e}=8\times10^{-4}$) and CPL model (with $w_0=-0.954$, $w_a=-0.29$) respectively. The red line in both panels shows the behaviour in standard $\Lambda$CDM cosmology.}
 \label{compareEDECPL}
\end{figure}

\capfigref{compareEDECPL}shows the equation of state as well as the matter and dark energy fractional densities for the EDE and CPL models. It is  clear that the two parameterizations have very different behaviour in the early epoch limit. In CPL models, $w$ is not constant but the dark energy density tends to zero when $a$ tends to zero. In contrast, the EDE models have $w(a) \rightarrow 0$ and the dark energy density is finite when $a \rightarrow 0$. The CPL parameterization corresponds to a Taylor expansion around $a=1$. Performing a similar expansion for the EDE equation of state gives 
\beq
 w(a)= \frac{w_0}{[1-b\ln [1-(1-a)]]^2} \sim \frac{w_0}{[1+b(1-a)]^2}  \sim w_0-2bw_0(1-a) + \mathcal{O}((1-a)^2)
 \label{Approximation}
\eeq
 Comparing with \eqnref{EoSCPL2}, we can relate $w_a=-2bw_0$, since $w_0$ denotes the same quantity in both expressions. For our choice of EDE parameters, we have $-2 b w_0 =0.73 $. Thus, although, \eqnref{EoSCPL2}  and \eqnref{Approximation} and  have the same form, the value of the parameters are different if consider the recent constraints from the observational data. In particular, the signs are different and the equation $w$ is less (EDE) and more (CPL) negative as compared to the $\Lambda$CDM value of -1. 
\section{Converting to the Eulerian PDF}
\label{app:lagtoeul}
Given a discrete distribution of matter with mean density ${\bar \rho}$, there are two ways to define the density field. Fix the volume $V_0$ around each (fixed grid) point and estimate the matter enclosed inside it to define the overdensity $\delta$. Alternately, fix the mass $M_0$ associated with each point and estimate the volume it occupies. The former corresponds to the Eulerian prescription whereas the latter to the Lagrangian prescription.  The fixed mass $M_0$ is related to the scale $R_f$ through 
\beq 
M_0 = \frac{4 \pi}{3} R_f^3 {\bar \rho}, 
\eeq
and the $\sigma_G$ is related to the scale $R_f$ through 
\beq 
\sigma^2(R_f) = \int P(k) W^2(k R_f) d^3 k, 
\eeq
where $P(k)$ is the power spectrum. The Eulerian and Lagrangian PDFs are related as 
\beq
(1+\delta)P_E(\delta, \sigma_G)=\left\{P_L(\delta,\sigma_G)-\frac{d\sigma_G}{dR_f}\frac{dR_f}{d\delta}\int_\delta^\infty\frac{dP_L(\delta',\sigma_G)}{d\sigma_G} d \delta'\right\}
\label{LtoE}
\eeq
This expression is equivalent to one derived by \citet{bernardeau_effects_1994}, equation 33, and has been re-derived in the context of local dynamics by \cite{nadkarni-ghosh_phase_2016}, appendix E. The PDF obtained by the method based on spherical collapse is Lagrangian since it assumes that the mass of each sphere in the 
ensemble is the same (set by the scale $\sigma_G$) and it remains constant throughout the evolution. However, the fitting forms based on Eulerian perturbation theory  or numerical simulations correspond to the Eulerian PDF. The Lagrangian PDF was converted to the Eulerian one using the above formula. However, we do not know $P_L(\delta',\sigma_G)$ analytically. Thus, to find the derivative of $P_L$ in the integrand,  we performed numerical runs at $\sigma_G=0.6, 1.1$ and $2.1$. The data for all $\sigma_G$ was binned in the same range and discrete derivatives were computed in each $\delta$ bin. The definition $\delta = M_0/{\bar \rho} V_0 -1$ gives the pre-factor $\frac{dR_f}{d\delta} = R_f/(1+\delta)$ (for fixed $V_0$). 
The initial realization is simply an ensemble of points drawn from a Gaussian and does not correspond to any power spectrum (the spherical dynamics assumes only local collapse; there is no reference to spatial correlations). Ideally, to compute the pre-factor $d \sigma_G/d R_F$, one needs to know the initial power spectrum that is chosen by a particular simulation, whose results one wants to compare. However, in this paper, we do not aim to compare our results to a particular output of a simulation. It is important to keep in mind that the forms given below only provide the analytic dependence on $\delta$. The parameters of the fitting functions (for e.g., $\sigma_\phi$ in SLN or $\alpha$, $\kappa$ and $\xi$ of $N_{v2}$) are computed from the data generated by spherical dynamics and not the N-body simulation and we only aim to check which form fits best. In view of this, we approximate the pre-factor $d \sigma_G/d R_F$  using a BBKS power spectrum under the assumption that the order of magnitude of the pre-factor will not be significantly different with any other more realistic power spectrum. The transformed PDF does not satisfy $\int P_E(\delta) d\delta = 1$; it is explicitly normalized by dividing by the norm. A more accurate transformation for general PDFs based on 1d systems is given in \citet{pajer_divergence_2018}.

\section{Best fit parameters for the $\mathcal{N}_{v2}$ distribution for the three cosmologies.  }
\label{app:bestfitparams}
\begin{table}
\smallskip
\begin{tabular*}{\textwidth}{@{\extracolsep{\fill}}|ccc|c|c|c|c|c|c|}
\hline
\multirow{2}{*}{Model} & \multirow{2}{*}{$\sigma_G$} & \multirow{2}{*}{$a$} & \multicolumn{2}{c|}{$\alpha$} & \multicolumn{2}{c|}{$\kappa$} & \multicolumn{2}{c|}{$\xi$} \\
\cline{4-9}
& & & Lagrangian & Eulerian & Lagrangian & Eulerian & Lagrangian & Eulerian \\
\hline
\multirow{3}{*}{$\Lambda$CDM} & \multirow{3}{*}{0.5} & 0.1 & 0.064 & 0.064 & -0.108 & -0.113 & 0.999 & 0.997   \\
& & 0.4 & 0.254 & 0.234 & -0.389 & -0.379 & 1.000 & 0.958   \\
 & & 1.0 & 0.504 & 0.351 & -0.670 & -0.599 & 0.988 & 0.856    \\[2ex]
 & \multirow{3}{*}{1.0} & 0.1 & 0.128 & 0.125 & -0.201 & -0.200 & 1.001 & 0.987  \\
 & & 0.4 & 0.476 & 0.342 & -0.645 & -0.586 & 0.993 & 0.824  \\
  & & 1.0 & 0.832 & 0.323 & -0.948 & -0.749 & 0.990 & 0.567    \\[2ex]
 & \multirow{3}{*}{2.0} & 0.1 & 0.256 & 0.230 & -0.389 & -0.365 & 0.9998 & 0.945  \\
 & & 0.4 & 0.766 & 0.305 & -0.987 & -0.716 & 0.904 & 0.543   \\
 & & 1.0 & 0.848 & 0.272 & -1.223 & -0.936 & 0.743 & 0.346    \\
\hline
\multirow{9}{*}{EDE} & \multirow{3}{*}{0.5} & 0.1 & 0.067 & 0.067 & -0.113 & -0.119 & 0.999 & 0.997   \\
 & & 0.4 & 0.245 & 0.228 & -0.377 & -0.372 & 1.000 & 0.962   \\
 & & 1.0 & 0.539 & 0.345 & -0.671 & -0.583 & 1.078 & 0.870    \\[2ex]
 & \multirow{3}{*}{1.0} & 0.1 & 0.135 & 0.132 & -0.213 & -0.210 & 1.001 & 0.986  \\
 & & 0.4 & 0.495 & 0.342 & -0.665 & -0.579 & 1.008 & 0.838  \\
  & & 1.0 & 0.728 & 0.331 & -0.867 & -0.742 & 0.977 & 0.592    \\[2ex]
 & \multirow{3}{*}{2.0} & 0.1 & 0.269 & 0.240 & -0.406 & -0.384 & 0.999 & 0.939  \\
 & & 0.4 & 0.612 & 0.307 & -0.840 & -0.707 & 0.863 & 0.549   \\
 & & 1.0 & 0.851 & 0.233 & -1.134 & -0.874 & 0.812 & 0.327    \\
\hline
\end{tabular*}
\caption{Values of the parameters $\alpha$, $\kappa$ and $\xi$ for best fitted $\mathcal{N}_{v2}$ distribution:}
\label{table:bestfit}
\end{table}

%

\section{Width of the one-point PDF for higher values of $\sigma_G$.}
\label{app:width}

\begin{figure}
\begin{center}
\includegraphics[width=15cm]{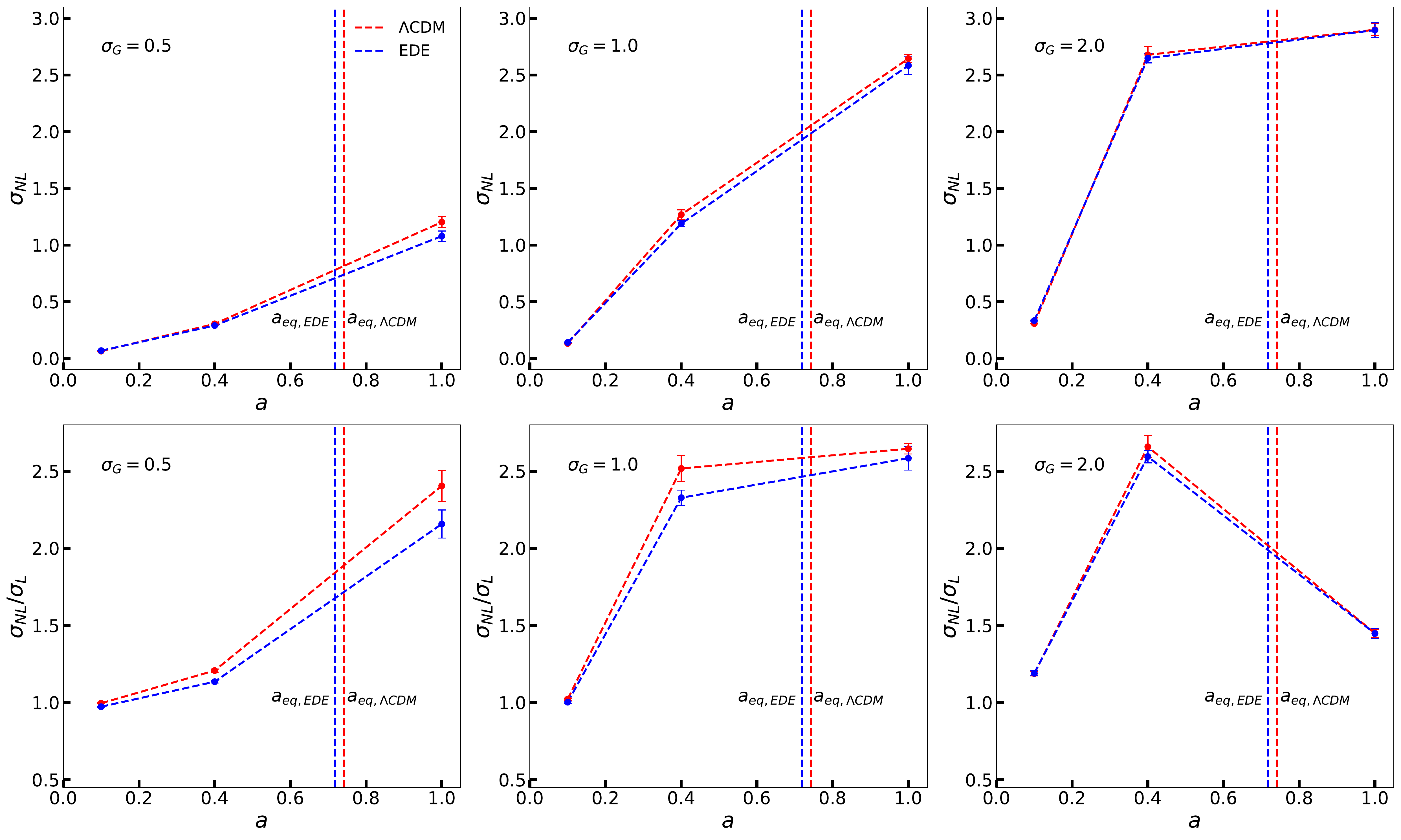}
\label{sigmaNLcoarse}
\caption{Width of the PDF (top panel) and scaled width (bottom panel) as a function of time.  The flattening for $\sigma_G=2$ is an effect of a finite cutoff imposed on the $\delta$ values in computing the PDF.}
\end{center}
\end{figure}

 The upper panel of \figref{sigmaNLcoarse} shows the variation of $\sigma_{NL}$ as a function of $a$ ($a=0.1,0.4,1$).  
It is clear that for all values of $\sigma_G$ the width of the PDF increases with time. Between $a=0.1$ and $a=0.4$, the rate of increase is higher for higher $\sigma_G$, which is expected because the non-linear growth rate is higher for higher $\delta$ values. From $a=0.4$ to $a=1$, there is an apparent slow-down in the growth rate for $\sigma_G=1,2$. While slow down is expected at higher epochs due to the presence of dark energy, the one observed slow down is an artefact of the cut-off in $\delta$ imposed in the construction of the PDF. For $\sigma_G=2$, that the change (increase) in $\sigma_{NL}$ from $a=0.4$ to $a=1$ was higher when the cut-off $\delta$ was increased from 5 (plot not shown) to 20 (this plot). 

%
\newpage
\bibliographystyle{mn2e}
\bibliography{EDE,book}

\label{lastpage}
\end{document}